\documentclass[%
reprint,
superscriptaddress,
nofootinbib,
amsmath,amssymb,
aps,
pra,
noeprint,
longbibliography
]{revtex4-2}

\usepackage{graphicx}
\usepackage[english]{babel}
\usepackage{appendix}
\usepackage{dcolumn}
\usepackage{bm}
\usepackage{color}
\usepackage{booktabs,eqparbox}
\usepackage{braket}
\usepackage{dsfont}
\usepackage{placeins}
\usepackage{paralist}
\usepackage{comment}
\usepackage[utf8]{inputenc}
\usepackage{pgfplots}
\usepackage{amsmath}
\usepackage{graphicx}
\usepackage{lipsum}
\usepackage{tikz}
\usepackage{csquotes}
\usetikzlibrary{shapes,arrows,fit, backgrounds, quantikz2}
\usepackage{hyperref}
\usepackage{cleveref}
\usepackage{textgreek}
\usepackage[printonlyused,withpage]{acronym}
\usepackage{quantikz}
\usepackage[version=4]{mhchem}
\usepackage{orcidlink}

\DeclareUnicodeCharacter{2212}{−}
\usepgfplotslibrary{groupplots,dateplot}
\usetikzlibrary{patterns,shapes.arrows}
\pgfplotsset{compat=newest}

\makeatletter
\let\oldtheequation\theequation
\renewcommand\tagform@[1]{\maketag@@@{\ignorespaces#1\unskip\@@italiccorr}}
\renewcommand\theequation{(\oldtheequation)}
\makeatother

\definecolor{myRed}{HTML}{A3061E}
\definecolor{myBlue} {RGB} {0,63,119}
\definecolor{myYellow} {cmy} {0,0.263,0.741}
\definecolor{myGreen}{HTML}{0B6E4F}
\definecolor{FAU}{RGB}{0,56,101}
\definecolor{gray75}{gray}{0.75}
\colorlet{myOrange} {myYellow!60!myRed}
\colorlet{myViolet}{myRed!50!myBlue!80}

\renewcommand{\vec}[1]{\bm{#1}}

\newcommand{\mycomment}[1]{}

\usetikzlibrary{quantikz2,fit,backgrounds,calc,positioning,arrows.meta}

\definecolor{gsfill}{RGB}{225,236,250}   
\definecolor{gsdraw}{RGB}{70,110,170}    
\definecolor{vfill}{RGB}{226,242,228}    
\definecolor{vdraw}{RGB}{70,140,90}      
\definecolor{dftc}{RGB}{90,90,90}
\definecolor{clc}{RGB}{52,101,164}
\definecolor{qc}{RGB}{56,142,60}
\definecolor{outc}{RGB}{200,110,30}

\tikzset{
  gsgate/.style={fill=gsfill, draw=gsdraw, thick},
  vgate/.style={fill=vfill, draw=vdraw, thick},
  >={Stealth[length=1.6mm]},
  box/.style={draw=#1, fill=#1!10, rounded corners=2pt, line width=0.6pt,
              align=center, font=\footnotesize, inner sep=3pt,
              minimum width=3.4cm, minimum height=9mm},
  lbl/.style={font=\scriptsize, text=black!65, align=center},
  arr/.style={->, line width=0.6pt, black!70},
}

\addto\extrasenglish{%
}

\addto\extrasenglish{%
}

\begin{document}

\title{Dynamical Mean-Field Theory for Real Materials using Variational Quantum Compressed Time Evolution on a Quantum Computer}

\author{Stefan Wolf \orcidlink{0009-0003-8297-3745}}
\affiliation{Department of Physics, Friedrich-Alexander-Universität Erlangen-Nürnberg, Erlangen, Germany}

\author{Johannes Selisko \orcidlink{0000-0001-6733-0782}}
\affiliation{Corporate Sector Research and Advance Engineering, Robert Bosch GmbH, Robert-Bosch-Campus 1, D-71272 Renningen, Germany}

\author{Timo Eckstein \orcidlink{0000-0002-6819-1865}}
\affiliation{Department of Physics, Friedrich-Alexander-Universität Erlangen-Nürnberg, Erlangen, Germany}

\author{Ludwig Nützel \orcidlink{0000-0002-4241-1241}}
\affiliation{Department of Physics, Friedrich-Alexander-Universität Erlangen-Nürnberg, Erlangen, Germany}

\author{Maximilian Amsler~\orcidlink{0000-0001-8350-2476}}
\affiliation{Corporate Sector Research and Advance Engineering, Robert Bosch GmbH, Robert-Bosch-Campus 1, D-71272 Renningen, Germany}

\author{Christopher Wever \orcidlink{0000-0003-4246-8464}}
\affiliation{Corporate Sector Research and Advance Engineering, Robert Bosch GmbH, Robert-Bosch-Campus 1, D-71272 Renningen, Germany}

\author{Georgy Samsonidze \orcidlink{0000-0002-3759-1794}}
\affiliation{Robert Bosch LLC, Research and Technology Center, Sunnyvale, CA 94085, United States of America}

\author{Thomas Eckl}
\affiliation{Corporate Sector Research and Advance Engineering, Robert Bosch GmbH, Robert-Bosch-Campus 1, D-71272 Renningen, Germany}

\author{Michael J. Hartmann \orcidlink{0000-0002-8207-3806}}
\affiliation{Department of Physics, Friedrich-Alexander-Universität Erlangen-Nürnberg, Erlangen, Germany}
\affiliation{Quint Computing GmbH, Henkestr. 91, 91052 Erlangen, Germany}

\date{\today}

    \begin{abstract}
       {
        We design a hybrid quantum-classical computational workflow, combining density functional theory and dynamical mean-field theory, to investigate the electronic properties of \ce{CaCuO2}, a parent structure of high-temperature cuprate superconductors.  
Our approach uses a quantum algorithm as an impurity solver that is based on variational quantum compressed time evolution to compute the time-dependent Green's function on an IBM Heron quantum device. Utilizing error mitigation techniques and a Lehmann representation fitting procedure, we  extract the density of states from the signal in the accessible time regime. 
Despite the significant challenges rooted in the residual imprecision of currently accessible NISQ devices, we successfully resolve the characteristic features of the spectral function. 
However, we find that the required numerical accuracy for full DMFT self-consistency exceeds current hardware capabilities, indicating that practical real-material DMFT applications will necessitate post-NISQ hardware developments.
       }
    \end{abstract}

\maketitle

\section{Introduction}
Strongly correlated materials represent a cornerstone of next-generation industrial and technological innovation. 
Unlike conventional semiconductors and metals, these systems exhibit extraordinary sensitivity to external stimuli such as temperature, pressure, and electromagnetic fields, making them ideal candidates for a wide array of advanced applications. 
Most profoundly, strongly correlated materials hold the key to potentially uncovering a scalable, high-temperature superconductor \cite{Keimer}. 
Since conventional electron-phonon coupling mechanisms fail to explain the superconducting mechanism in oxide families such as cuprates or nickelates, decoding the strong electronic correlations within these materials is not merely an academic exercise but an industrial imperative for the future of technology \cite{Norman}. 

Treating the quantum many-body systems governing the properties of correlated materials using analytic tools is rendered impossible by the growth of entanglement and Hilbert space dimensions caused by the strong interaction between the electrons. 
Even standard first-principles computational methods, most notably Density Functional Theory (DFT), fundamentally struggle with these strong correlations \cite{Kotliar}. 
To properly capture the physics of the Mott state, the pseudogap regime, and the eventual transition into a superconducting phase, one must explicitly treat the competition between the kinetic hopping amplitude and the local Coulomb repulsion, as idealized by the Hubbard model \cite{Arovas}.

Dynamical Mean-Field Theory (DMFT) \cite{Georges} has emerged as a highly successful framework to solve such strongly interacting lattice models.
By mapping the intractable lattice problem onto a purely local yet fully interacting quantum impurity model embedded in a self-consistent fermionic bath, DMFT captures the critical temporal fluctuations and local many-body correlations that static mean-field theories miss. 
To bridge the gap between idealized model Hamiltonians and realistic structures, the synthesis of DFT and DMFT, denoted as DFT+DMFT, has become a powerful methodology for modeling the behavior of correlated electron systems. 
In this hybrid approach, DFT provides the precise single-particle band structure and hybridization functions, while DMFT rigorously treats the strong local interactions within the correlated orbital subspace. 

Nevertheless, DMFT still requires solving a quantum many-body problem, from the easiest case with a single impurity like the single-impurity Anderson model (SIAM) to the tougher challenge with a larger number of impurities (cluster DMFT). 
The application of state-of-the-art classical impurity solvers based on the Numerical Renormalization Group (NRG) and the Density Matrix Renormalization Group (DMRG) \cite{WolfCheby, Stadler, Kugler, Bauernfeind2017,Bauernfeind2018} is limited by the exponential growth of the Hilbert space with the number of bath orbitals, especially in the case of cluster impurities.
Quantum Monte Carlo (QMC) \cite{Gull}, a method that circumvents this issue, suffers instead from the fermionic sign problem and does not give insight into spectral information. 

Solving the impurity model is, therefore, a prime candidate for quantum computing to overcome these limitations.
In this hybrid quantum-classical approach, the quantum device is used to compute the Green's function of the impurity sites.
After preparing the ground state of the impurity model on the quantum computer, the Green's function can be either reconstructed by measuring different observables \cite{Endo, Rungger, Jamet, Rizzo, Ehrlich, Selisko} or directly measured~\mbox{\cite{Bauer, Kreula, Jaderberg, Keen, Steckmann, Hogan}}.

In this work, we use the latter approach in a hybrid quantum-classical DFT+DMFT framework to investigate the electronic properties of the strongly correlated material \ce{CaCuO2}.
We perform a variational quantum compression of the time evolution operator \cite{Jaderberg, Benedetti, Barison, Berthusen, Puig, Goh, Wolf} into a Hamiltonian variational ansatz with a fixed circuit depth, significantly reducing the depth, such that it can run on the IBM Heron device (ibm\_torino).
After performing the quantum measurements, we apply targeted error mitigation and classical post-processing to extract the density of states.
Despite the challenges associated with NISQ devices, we show that this method successfully captures the broader system dynamics and resolves the critical spectral features of \ce{CaCuO2}, including the lower and upper Hubbard bands as well as the central quasiparticle peak.
Alongside these material-specific results, we present a thorough analysis of the current hardware limitations, detailing how dephasing-induced damping and device calibration drift affect the measurements of the impurity Green's function. 
Consequently, we discuss the implications of these hardware constraints, concluding that, while qualitative spectral analysis is currently viable, achieving the strict numerical accuracy required for fully self-consistent DMFT loops is not feasible on current NISQ hardware.

\section{Method}

\label{sec:method}
Cuprates are an important class of high-temperature superconducting materials.
Superconductors in this category consist mainly of copper-oxygen planes with a square-lattice structure, separated by electronically insulating layers, e. g., calcium (\ce{Ca}) and other structures that act as spacer or charge reservoir layers. 
The underlying infinite-layer structure of the \ce{CaCuO2} compound is therefore often termed the 'parent structure' of high-temperature superconductors.
Hence, it is of particular interest to understand the electronic properties of \ce{CaCuO2} in the search for stable high-temperature superconductors under ambient conditions.
This setup provides an ideal test case for the workflow introduced in this section, combining DFT and DMFT calculations.

\subsection{DFT+DMFT Workflow}
\label{subsec:dftdmft}
We employ the DFT+DMFT workflow as presented by Selisko~\textit{et al.}~\cite{Selisko}, utilizing the solid\_dmft framework~\cite{Merkel2022,beck_charge_2022}. 
For detailed workflow insights, the reader is referred to the original publications and references therein. 
The following paragraphs explain the details of the \ce{CaCuO2} calculations performed in this work, which are summarized in Fig. \ref{fig:workflow}.

Our procedure starts with a self-consistent DFT calculation via the Quantum ESPRESSO package~\cite{giannozzi_quantum_2009}, employing the Perdew-Burke-Ernzerhof (PBE) exchange-correlation functional~\cite{perdew_generalized_1996} alongside norm-conserving pseudopotentials~\cite{van_setten_pseudodojo_2018}.
A plane wave cutoff energy of 100 Ry (Ry $\approx$ 13.6 eV) and a $12\times12\times12$ k-point mesh ensure a properly converged electronic structure. 
We construct a tight-binding model of the $d_{x^2-y^2}$-band, intersecting the Fermi level, using Wannier90~\cite{pizzi_wannier90_2020}, with energy windows set from $2.00\,\text{e}V$ to $16.00\,\text{eV}$ (outer) and from $10.00\,\text{eV}$ to $11.30\,\text{eV}$ (inner), centered around the Fermi level at $10.28\, \text{eV}$. The wan2respack~\cite{kurita_interface_2023} tool and RESPACK's cRPA implementation~\cite{nakamura-20-respack} result in a Hubbard $U$ value of $3.11\,\text{eV}$, aligning with findings from Ref.~\cite{Karp_CaCuO2}.

In our study, we execute one-shot DMFT calculations, omitting the feedback of charge corrections into DFT. 
For an extensive examination of \ce{CaCuO2} that includes cluster DMFT and antiferromagnetic studies, see Ref.~\cite{Karp_CaCuO2_SC}. 
Our procedure consists of three steps:

\begin{enumerate}
    \item We conduct a self-consistent DMFT analysis utilizing the continuous-time hybridization expansion (CTHYB) solver, integrated within TRIQS~\cite{seth_triqscthyb_2016,triqs}. This involves executing $10^8$ sweeps with a cycle length of 400 at an inverse temperature of $\beta=30\,\text{eV}^{-1}$.
    \item Based on the converged results of the CTHYB solver, we proceed with a self-consistent DMFT computation employing exact diagonalization (ED) at $0\,\text{K}$. This step utilizes the identical imaginary frequency grid as the CTHYB calculation, despite the temperature difference.
    \item  The final ED-DMFT iteration is recomputed using the quantum computing impurity solver of Ref.~\cite{Wolf}.
\end{enumerate}
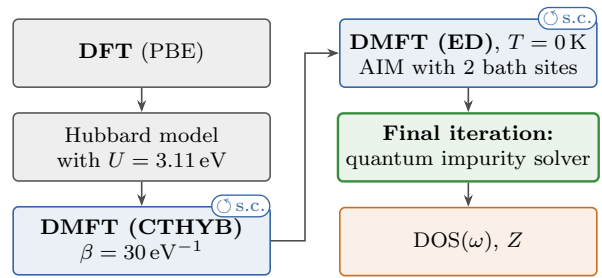
\begin{figure}
\begin{tikzpicture}[node distance=3.2mm]

\node[box=dftc] (dft) {\textbf{DFT} (PBE)};
\node[box=dftc, below=of dft] (wan) {Hubbard model\\ with $U = 3.11$\,eV};
\node[box=clc,  below=of wan] (cthyb) {\textbf{DMFT (CTHYB)}\\ $\beta = 30\,\mathrm{eV}^{-1}$};

\node[box=clc,  right=9mm of dft] (ed) {\textbf{DMFT (ED)}, $T=0\,\text{K}$\\ AIM with 2 bath sites};
\node[box=qc,   below=of ed, line width=0.9pt] (qs) {\textbf{Final iteration:}\\ quantum impurity solver};
\node[box=outc, below=of qs] (out) {DOS$(\omega)$, $Z$};

\foreach \a/\b in {dft/wan,wan/cthyb,ed/qs,qs/out} \draw[arr] (\a) -- (\b);

\coordinate (mid) at ($(cthyb.east)!0.5!(ed.west)$);
\draw[arr] (cthyb.east) -- (cthyb.east -| mid) -- (ed.west -| mid) -- (ed.west);

\tikzset{badge/.style={draw=clc, fill=white, rounded corners=4pt, line width=0.5pt,
         font=\scriptsize, text=clc, inner sep=1.5pt}}
\foreach \n in {cthyb,ed}
  \node[badge] at ([xshift=-4mm]\n.north east) {$\circlearrowleft$\,s.c.};
\end{tikzpicture}

\caption{DFT+DMFT workflow for \ce{CaCuO2}. A DFT calculation (PBE) is downfolded onto a single-band Hubbard model with interaction strength $U = 3.11\,\text{eV}$. The DMFT equations are first converged with a CTHYB solver at $\beta = 30\, \text{eV}^{-1}$ and subsequently with an ED solver at $T = 0\,\text{K}$, using an Anderson impurity model with two bath sites ($\circlearrowleft$\, s.c.\ denotes self-consistent DMFT loops). The final ED-DMFT iteration is recomputed with the quantum impurity solver to obtain the density of states $\text{DOS}(\omega)$ and the quasiparticle weight $Z$.} 
\label{fig:workflow}
\end{figure}

In the ED calculation, we construct an Anderson impurity model (AIM) featuring two bath sites. 
This is achieved by fitting a Lehmann representation (see Eq. \ref{eq:Lehmann_rep}) to the hybridization function $\Delta(i\omega_n)$. 
Employing the same approach as detailed in Ref.~\cite{Selisko}, we derive the linear AIM characterized by the Hamiltonian (in eV):
\begin{equation}
\begin{split}
    H = &Un_{0,\uparrow}n_{0,\downarrow} + \sum_{p=0,\sigma}^2 \epsilon_pn_{p,\sigma}\sum_{p = 0,\sigma}^1 + V_p[c_{p,\sigma}c_{p+1,\sigma}^\dag + h.c.]\\
    = &3.11n_{0,\uparrow}n_{0,\downarrow}  -\sum_\sigma\left[1.31n_{0,\sigma} + 0.08n_{1,\sigma} -0.01n_{2,\sigma}\right]\\
    &+ \sum_\sigma \left[0.51(c_{0,\sigma}c^\dagger_{1,\sigma}+h.c.)+ 0.16(c_{1,\sigma}c^\dagger_{2,\sigma}+h.c.)\right],
\label{eq:SIAM}
\end{split}
\end{equation}
where index 0 denotes the impurity site and indices 1 and 2 correspond to the two bath sites. 
The term $+h.c.$ indicates the inclusion of the Hermitian conjugate and the operators $c, c^\dagger, n$ are the fermionic annihilation, creation, and number operators, respectively.

The ED solver and the quantum computing solver yield a Lehmann representation of the zero-temperature retarded Green's function,
\begin{equation}
    G(z) = \sum_j \frac{\lambda_j}{z-\omega_j}.\label{eq:Lehmann_rep}
\end{equation}
In the ED method, the position $\omega_j$ of the peak $j$ is identified as the energy difference between the ground state and the $j$th excited state, with an additional negative sign if state $j$ contains fewer electrons than the ground state. 
The amplitude 
\begin{equation}
    \lambda_j=|\bra{j}c_{0,\sigma}+c_{0,\sigma}^\dagger\ket{\text{GS}}|^2
    \label{eq:amplitude}
\end{equation}
is derived from the eigenstates of the Hamiltonian. 
The complex variable $z$ may represent Matsubara frequencies $i\omega_n$ to continue the DMFT computation or real frequencies with a minor imaginary part $\omega+i\eta$ to calculate the final real quantities of interest.

The real-frequency Dyson equation requires careful handling of numerical artifacts arising from the Green's function's sharp peaks. We adopt the pole representation strategy for storing the Green's function as suggested in Refs.~\cite{Selisko, Lu_Efficient}. 
The CTHYB density of states is obtained using maximum entropy analytic continuation.

The quasiparticle weight, defined via the self-energy's derivative, is numerically unstable. 
Minor peaks, potentially introduced by quantum hardware noise, can significantly change the derivative.
Lu~\textit{et al.} \cite{Lu_Efficient} and Selisko~\textit{et al.} \cite{Selisko} propose a regularization technique to redistribute peaks near the Fermi level. 
In this work, we define the quasiparticle weight as the integral over the quasiparticle peak of the density of states, with integral boundaries determined by the lowest local minima below and above the Fermi level, which we find to be numerically more stable for discrete spectra.

\subsection{Impurity Solver}
\label{subsec:impuritysolver}
The intention is to perform the final ED-DMFT iteration on a quantum device following the scheme presented in Ref. \cite{Wolf}.
This includes the following steps:
\begin{enumerate}
    \item prepare the ground state $\ket{\text{GS}} = U_\text{GS}\ket{0}$ with a variational quantum eigensolver (VQE) \cite{Peruzzo}
    \item use a variational quantum algorithm (VQA) to compress the Trotterized time evolution into a Hamiltonian variational ansatz with a fixed circuit depth
    \item measure the time-dependent Green's function $G(t)$ on the quantum device at discrete time steps $n\Delta t$
    \item perform error mitigation and classical post-processing to find the Lehmann parameters required for the next iteration step
\end{enumerate}
The Green's function of the impurity system in time is given by
\begin{equation}
    G(t) = -i\theta(t)\left(\braket{\{c_{0,\sigma}(t),c^\dag_{0,\sigma}(0)\}}\right)
    \label{eq:timeGF}
\end{equation}
with the anticommutator $\{.,.\}$ and the expectation value calculated w. r. t. the ground state. 
The time-dependent ladder operators are given by $c_{0,\sigma}^{(\dag)} = U^\dag(t)c_{0,\sigma}^{(\dag)}U(t)$ with the time evolution operator $U(t)$.
Using the Jordan--Wigner transformation \cite{Jordan} (see Appendix \ref{app:vqtec} for more details), the Green's function reads
\begin{equation}
\begin{split}
    G(t) = &\frac{-i}{2}\left(\text{Re}(\braket{U^\dag(t) X_{0,\sigma}U(t)X_{0,\sigma}})\right.\\
    &+ \left.i\text{Re}(\braket{U^\dag(t) Y_{0,\sigma}U(t)X_{0,\sigma}})\right).
\end{split}
\label{eq:timeGFJWT}
\end{equation}
Both expectation values in Eq. \ref{eq:timeGFJWT} can be measured using a Hadamard test (see Fig. \ref{fig:HVA}). 
For the variational time evolution compression in step 2, a single Trotter step $U(\Delta t)$ is used to iteratively optimize a sequence of circuits $V(\vec{\theta}_n)$ such that the two relations $V(\vec{\theta}_n)(X_{0,\sigma})\ket{\text{GS}} \approx U(n\Delta t)(X_{0,\sigma})\ket{\text{GS}}$ are fulfilled.
It is important that the time evolution approximation evolves both states, $X_{0,\sigma}\ket{\text{GS}}$ and $\ket{\text{GS}}$, correctly, as our approximation acts on these states in Eq. \ref{eq:timeGFJWT}.
This is done by evolving the system over time using $U(\Delta t)V(\vec{\theta}_{n-1})$ and then optimizing $V^\dag(\vec{\theta}_n)$ so that it evolves the state back to the two initial states.
We define $L_n = V^\dag(\vec{\theta}_n)U(\Delta t)V(\vec{\theta}_{n-1})$ and $K_n = X_{0,\sigma}L_nX_{0,\sigma}$.
The cost function is then given by 
\begin{equation}
    C(\vec{\theta}_n) = \sum_{p,\sigma} \left(1 - \text{Re}(\braket{\text{GS}|L_n^\dag U_\text{GS}\left(\ket{0}\bra{0}\right)_{p,\sigma}U^\dag_\text{GS}K_n|\text{GS}})\right)
    \label{eq:costfunction}
\end{equation}
where the summation over $p$ and $\sigma$ runs over all system qubits.
To compute these expectation values, it is sufficient to apply $K_n$ only once (see Appendix \ref{app:vqtec}).

The Green's function Eq. \ref{eq:timeGFJWT} is phase sensitive. 
Training the parametrized ansatz using the cost function \eqref{eq:costfunction} not only leads to a correct approximation but also avoids a global phase difference, which would require additional measurements. 
For $V(\vec{\theta})$, we choose a Hamiltonian variational ansatz where one layer of the  ansatz is a parametrized version of the Trotter step $U(\Delta t)$ (see Fig. \ref{fig:HVA}).
This ansatz conserves the symmetries of the system and therefore leads to fewer errors than more general choices.
These ansatz symmetries can also be used to reduce the circuit depth by removing $U^\dag_\text{GS}$ required to evaluate some of the expectation values in the cost function Eq. \ref{eq:costfunction}.
However, despite these simplifications, it is not yet feasible to run the circuit training on current quantum hardware.
We therefore restrict ourselves to just evaluating the time-dependent Green's function $G(t)$ after obtaining the circuits for the ground state preparation and the time evolution classically.
To this end, the pretrained circuits are compiled into a hardware-native gate set to measure the two expectation values in Eq. \eqref{eq:timeGFJWT}. 
Afterwards, we perform a Fourier transformation,
\begin{equation}
    G(\omega+i\eta) \approx \int_0^{t_\text{max}}dt\, e^{-\eta t}e^{i\omega t}G(t)
    \label{eq:fourier}
\end{equation}
with $\eta = 0.1$ to obtain some of the Lehmann frequencies $\omega_j$, where $t_\text{max}$ denotes the maximal time for which the Green's function is computed. 
These frequencies are then used as an initial guess to fit the Lehmann representation to $G(t)$ and determine the corresponding amplitudes $\lambda_j$ (see Sec. \ref{subsec:postprocess}). 
The density of states of the system is then given by
\begin{equation}
    \text{DOS}(\omega) = -\frac{1}{\pi}\text{Im}(G(\omega+i\eta)).
\end{equation}
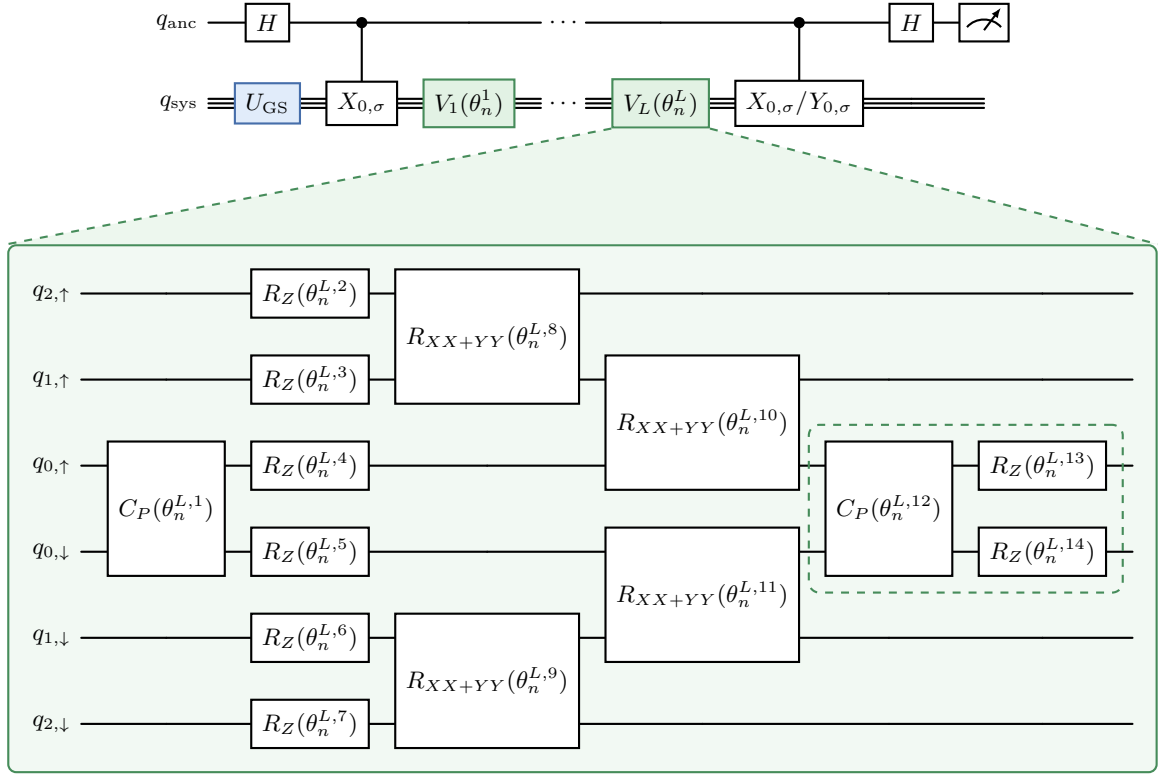
\begin{figure*}
\centering

\begin{quantikz}[remember picture, wire types={q,b}, classical gap=0.06cm, column sep=0.35cm, row sep=0.5cm]
\lstick{$q_\text{anc}$} & \gate{H} & \ctrl{1} &  & \ \cdots\ &  & \ctrl{1} & \gate{H} & \meter{}\\
\lstick{$q_\text{sys}$} & \gate[style=gsgate]{U_\text{GS}} & \gate{X_{0,\sigma}} & \gate[style=vgate]{V_1(\theta^1_n)} & \ \cdots\ & \gate[style={vgate,alias=VL}]{V_L(\theta^L_n)} & \gate{X_{0,\sigma}/Y_{0,\sigma}} &  & 
\end{quantikz}

\vspace{1.4cm}

\begin{quantikz}[remember picture, column sep=0.35cm,
  execute at end picture={
    \begin{pgfonlayer}{background}
      \node[fit=(current bounding box), inner sep=6pt, rounded corners=3pt,
            fill=vfill!45, draw=vdraw, thick] (zoombox) {};
    \end{pgfonlayer}}]
\lstick{$q_{2,\uparrow}$} &  & \gate{R_Z(\theta_n^{L,2})} & \gate[2]{R_{XX+YY}(\theta_n^{L,8})} &  &  &  & \\
\lstick{$q_{1,\uparrow}$} &  & \gate{R_Z(\theta_n^{L,3})} &  & \gate[2]{R_{XX+YY}(\theta_n^{L,10})} &  &  & \\
\lstick{$q_{0,\uparrow}$} & \gate[2]{C_P(\theta_n^{L,1})} & \gate{R_Z(\theta_n^{L,4})} &  &  & \gate[2]{C_P(\theta_n^{L,12})}\gategroup[2,steps=2,style={dashed,rounded corners,draw=vdraw,thick,inner xsep=3pt,inner ysep=3pt}]{} & \gate{R_Z(\theta_n^{L,13})} & \\
\lstick{$q_{0,\downarrow}$} &  & \gate{R_Z(\theta_n^{L,5})} &  & \gate[2]{R_{XX+YY}(\theta_n^{L,11})} &  & \gate{R_Z(\theta_n^{L,14})} & \\
\lstick{$q_{1,\downarrow}$} &  & \gate{R_Z(\theta_n^{L,6})} & \gate[2]{R_{XX+YY}(\theta_n^{L,9})} &  &  &  & \\
\lstick{$q_{2,\downarrow}$} &  & \gate{R_Z(\theta_n^{L,7})} &  &  &  &  &
\end{quantikz}

\begin{tikzpicture}[remember picture, overlay]
  \fill[vfill, opacity=0.6]
    (VL.south west) -- (VL.south east) -- (zoombox.north east) -- (zoombox.north west) -- cycle;
  \draw[vdraw, thick, dashed] (VL.south west) -- (zoombox.north west);
  \draw[vdraw, thick, dashed] (VL.south east) -- (zoombox.north east);
\end{tikzpicture}

\caption{Hadamard test for measuring the Green's function $G(t)$ using the approximation $V(\vec{\theta}_n)=V_L(\theta_n^L)\dots V_1(\theta^1_n)$ with $L$ layers of the Hamiltonian variational ansatz. The final layer is followed by an additional $C_P$-gate and $R_Z$-gates acting on the impurity sites (as illustrated here). The circuit for the ground state preparation $U_\text{GS}$ is obtained via VQE and a Hamiltonian variational ansatz.}
\label{fig:HVA}
\end{figure*}

\subsection{Error Mitigation}
\label{subsec:errormitigation}
Due to the Fourier transformation of $G(t)$ and the following fitting procedure, the impurity solver is resilient to small errors of single points of $G(t)$ as long as its shape is qualitatively well captured. 
As mentioned before, we expect damping effects caused by the depth of the circuit for ground state preparation and the increasing circuit depth dependent on the number of layers $L$. 
These can be modeled as
\begin{equation}
    G(t) = e^{\alpha_L \cdot L + \alpha_\text{GS}} G_L(t),
\end{equation}
where $G_L(t)$ denotes the measured Green's function with $L$ layers, for the first points until their curves diverge.
To verify this relation between damping and circuit depth, we trained the same parametrized ansatz for the case where the interaction between spin-up and spin-down is set to zero in the Hamiltonian Eq. \ref{eq:SIAM}, i. e., $U$ is set to zero.
We refer to this Hamiltonian as $H^0$.
The exact value of the Green's function $G^0(t)$ for $H^0$ but with respect to the ground state of $H$ can be calculated by performing the Fourier transformation of
\begin{equation}
\begin{split}
    G^{0}(\omega + i\eta) &=  \left(\omega + i\eta + 1.31 \right.\\
    &\left.- \frac{0.51^2}{\omega+i\eta + 0.08} - \frac{0.16^2}{\omega + i\eta - 0.01}\right)^{-1},
\end{split}
\label{eq:freeGF}
\end{equation}
which can be efficiently done on classical hardware.
In a Trotter approximation of the time evolution operator generated by $H^0$, no entanglement between the spin-up and spin-down sectors would build up, other than the amount already generated in the ground state preparation. 
Yet the optimization of the ansatz results in a very slight interaction between these sectors since the parameters of the $C_P$-gates acting on the qubit representing the impurity for spin-up and spin-down (see Fig. \ref{fig:HVA}) are not zero. 
The circuits for computing  $G^{0}(t)$ would still not generate significant entanglement between the spin sectors, but they have the same depth as the ones for $G(t)$ and the resulting damping caused by the depth is comparable.
Since we can compute the analytic expression Eq. \ref{eq:freeGF}, we can use these results to obtain the parameters $\alpha_L$ and $\alpha_\text{GS}$.

However, in practice, this procedure would result in a high cost in resources since additional circuits must be optimized and then evaluated, both using quantum hardware.
Alternatively, we can utilize the fact that the real component of the Green's function must be equal to one for $t=0$. 
Since the approximation of the time evolution operator reduces to the identity and damping is invisible for $t=0$, we instead extrapolate the measured values of the Green's function for the first points in time at $t = 0.08$ and $t = 0.16$ back to $t=0$. 
Nevertheless, this error mitigation only allows for a correction of the amplitude of $G(t)$ for smaller times.
As the entanglement grows over time, dephasing errors destroy more information, such that the damping effect for later times is stronger. 
We therefore expect this property to appear for $G(t)$ but not for $G^{0}(t)$ and measure both circuits to confirm this behavior.
Furthermore, this error mitigation cannot compensate for errors in the frequencies obtained by the Fourier transformation, but still improves the performance of the fitting scheme to find additional Lehmann parameters.

\subsection{Postprocessing of \protect\boldmath${G(t)}$}
\label{subsec:postprocess}
We follow the approach of Rungger ~\textit{et al.} and Wolf ~\textit{et al.} \cite{Rungger, Wolf} and use the Lehmann representation, Eq. \ref{eq:Lehmann_rep}, as a fitting function to extract as many poles as possible. 
Since we are restricted to a short time range, only a few of the $\lambda_j$ and $\omega_j$ can be determined. 
We therefore use the conditions
\begin{align}
    \sum_j \lambda_j &= 1\label{LPconditionSum}\\
    G(\omega = \epsilon_p) &= 0 \quad \forall p\label{LPconditionRoot}\\
    \left.\frac{\partial G(\omega)}{\partial \omega}\right|_{\omega = \epsilon_p} &= \frac{1}{V_p^2} \quad \forall p\label{LPconditionDeriv}
\end{align}
during the fitting procedure to include further Lehmann parameters.
Although these conditions are originally formulated for the parameters $V_p,\epsilon_p$ of the single-impurity Anderson model with a star topology, they can still be applied in our chain geometry while using the parameters of the star topology, since $G(t)$ is identical in both cases. 

The performance of the fitting procedure depends heavily on the initial guess. 
Intuitively, the Fourier transformation is a good starting point. 
However, performing it for the complete available time range yields only a small number of possible frequencies, such that the initial guess for additional frequencies must be chosen at random.
To obtain more educated initial guesses, we perform the Fourier transformation as given in Eq. \eqref{eq:fourier} with an increased $\eta$ and apply a short-time Fourier transformation,
\begin{equation}
    G_\tau(\omega) = \int_0^{t_\text{max}} dt\,w(t - \tau) e^{i\omega t} G(t)
\end{equation}
with a window function $w(t)$ for different segments of the signal (see Appendix \ref{app:postprocessing}). 
In this way, we obtain different sets of Lehmann parameters for each window which are all used as an initial guess for the fitting procedure during which the $\lambda_j$ for incorrectly chosen frequencies are mapped close to zero in the best case.

\subsection{Quasiparticle Weight}
\label{subsec:qpw}
The quasiparticle weight is a measure of the conductivity of the material.
It is defined as
\begin{equation}
    \mathcal{Z} = \frac{1}{1 - \left.\frac{d\text{Re}(\Sigma(\omega+i\eta))}{d \omega}\right|_{\omega=0}}
    \label{eq:qpw}
\end{equation}
with the self-energy given by
\begin{equation}
    \Sigma(\omega) = \left(\omega - \sum_{p=1}^2 \frac{V_p^2}{\omega - \epsilon_p}\right)^{-1} - G^{-1}(\omega)
\end{equation}
and serves as a fundamental wavefunction renormalization factor that quantifies the strength of electronic correlations.
As we expect a noisy signal that renders the calculation of the derivative of $\Sigma(\omega)$ and therefore of $\mathcal{Z}$ impossible using the above definition,  we instead determine the value qualitatively by integrating over the quasiparticle peak (for details, see Appendix \ref{app:qpw}).

\section{Results}
\begin{figure*}[t]
    \centering
    \includegraphics[width=1\textwidth]{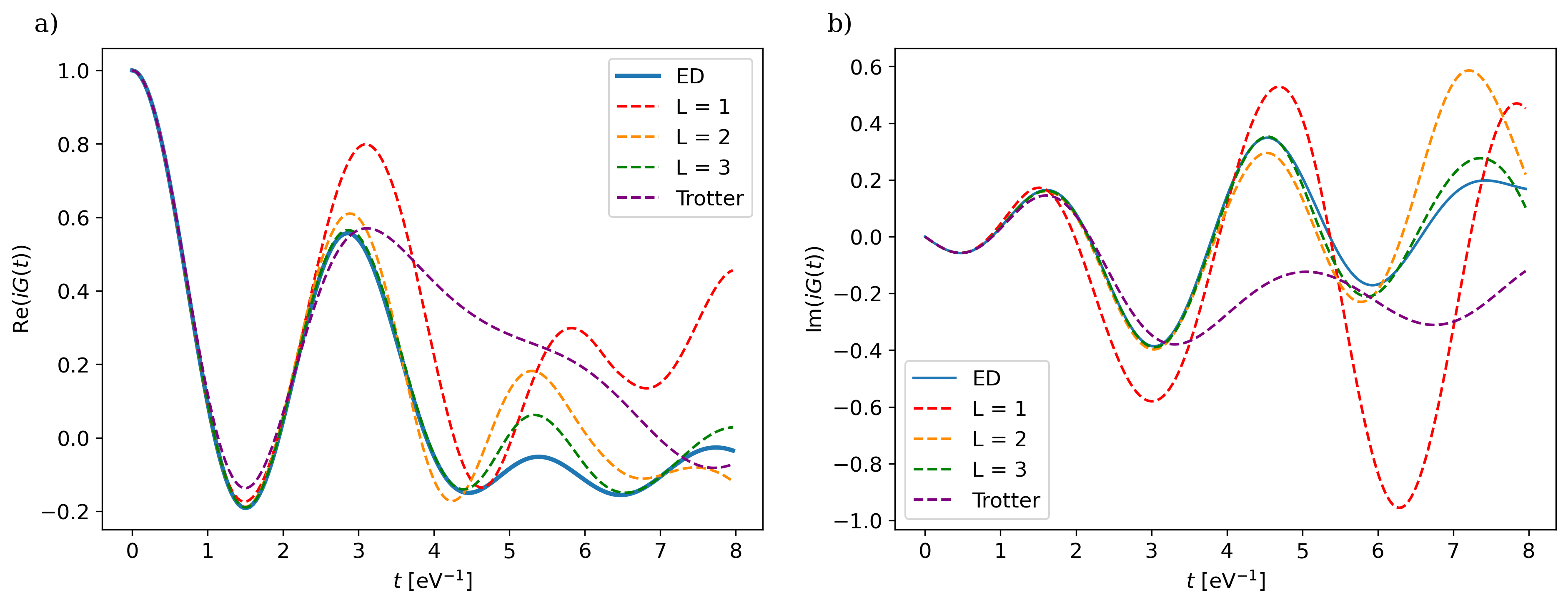}
    \caption{Real (a) and imaginary (b) components of $iG(t)$ obtained with the circuits using the VQA-Compression algorithm with $L\in\{1,2,3\}$ in a state-vector simulation. We compare our trained circuits to a first-order Trotter expansion with the same circuit depth as $L = 3$.}
    \label{fig:vqaResults}
\end{figure*}

We compare our optimized circuits with a Trotterization using the same gate count as for $L = 3$, i. e., for each $t$, the Trotterization is given by
\begin{equation}
    \hat{U}(t) = \left(\prod_j e^{-i\frac{H_jt}{3}}\right)^3 + \mathcal{O}(t^2)
\end{equation}
where $H_j$ are the terms of $H$ in Eq. \ref{eq:SIAM} grouped by their commutator behavior.
We refer to this approximation as a fixed Trotterization.
As illustrated in Fig. \ref{fig:vqaResults}, each parametrized circuit correctly approximates $G(t)$ only up to a certain time, beyond which the quantum circuit is no longer expressive enough.
The parametrized circuit for $L=1$ may quickly diverge from the actual value, but it follows the actual curvature better than the fixed Trotterization.
Circuits with $L=2$ and $L=3$ outperform our Trotterization with the same gate count also in terms of amplitude.
However, even with an optimized circuit with $L=3$, $G(t)$ is only approximated correctly until $t \approx 4$. 
This is expected, as for this number of bath sites, more gates are required to have an expressive enough circuit \cite{Wolf}.
In practice, one would actually not know for how long the Green's function is correctly calculated. 
We therefore chose to perform the calculation up to $t_\text{max} = 8$ and willingly include points that are not exactly correct to see the effects on the shape of the density of states.

Due to the small $t_\text{max}$, the Fourier transformation is only able to resolve four $\omega_j$.
Nevertheless, using the procedure described in Sec. \ref{subsec:postprocess}, we are able to extract a density of states for the impurity model that is in agreement with the one obtained from exact diagonalization, admittedly with a large broadening of $\eta = 0.1$ (see Appendix \ref{app:postprocessing} for details). 

\begin{figure*}[t]
    \includegraphics[width = 1\textwidth]{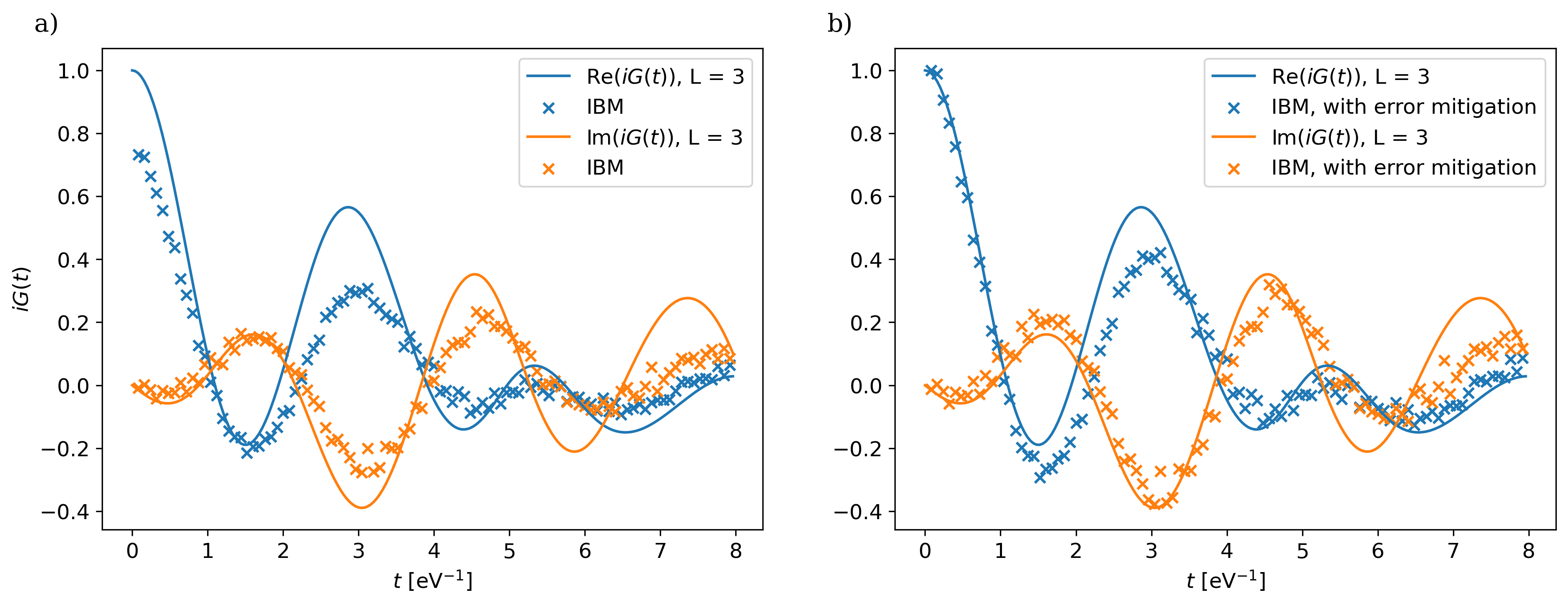}
    \caption{Real component of $iG(t)$ obtained with the pretrained circuits using 65 two-qubit gates on the IBM Heron quantum computer without (a) and with (b) error mitigation. The number of shots is 8192 for each point in time.}
    \label{fig:hardwareResults}
\end{figure*}

Having this insight, we aim to verify these results on a quantum computer. 
The target quantum device, the IBM Heron chip (ibm\_torino), uses a native gate set containing the $C_Z$-gate as the entangling gate.
After compiling, the trained circuits consist of 65 of these two-qubit gates, of which 35 are used to prepare the non-trivial ground state and the Hadamard test for measurement (for details, see Appendix \ref{app:qct}).
The results are illustrated in Fig. \ref{fig:hardwareResults}.
\begin{figure}[t]
    \centering
    \includegraphics[width = 1\linewidth]{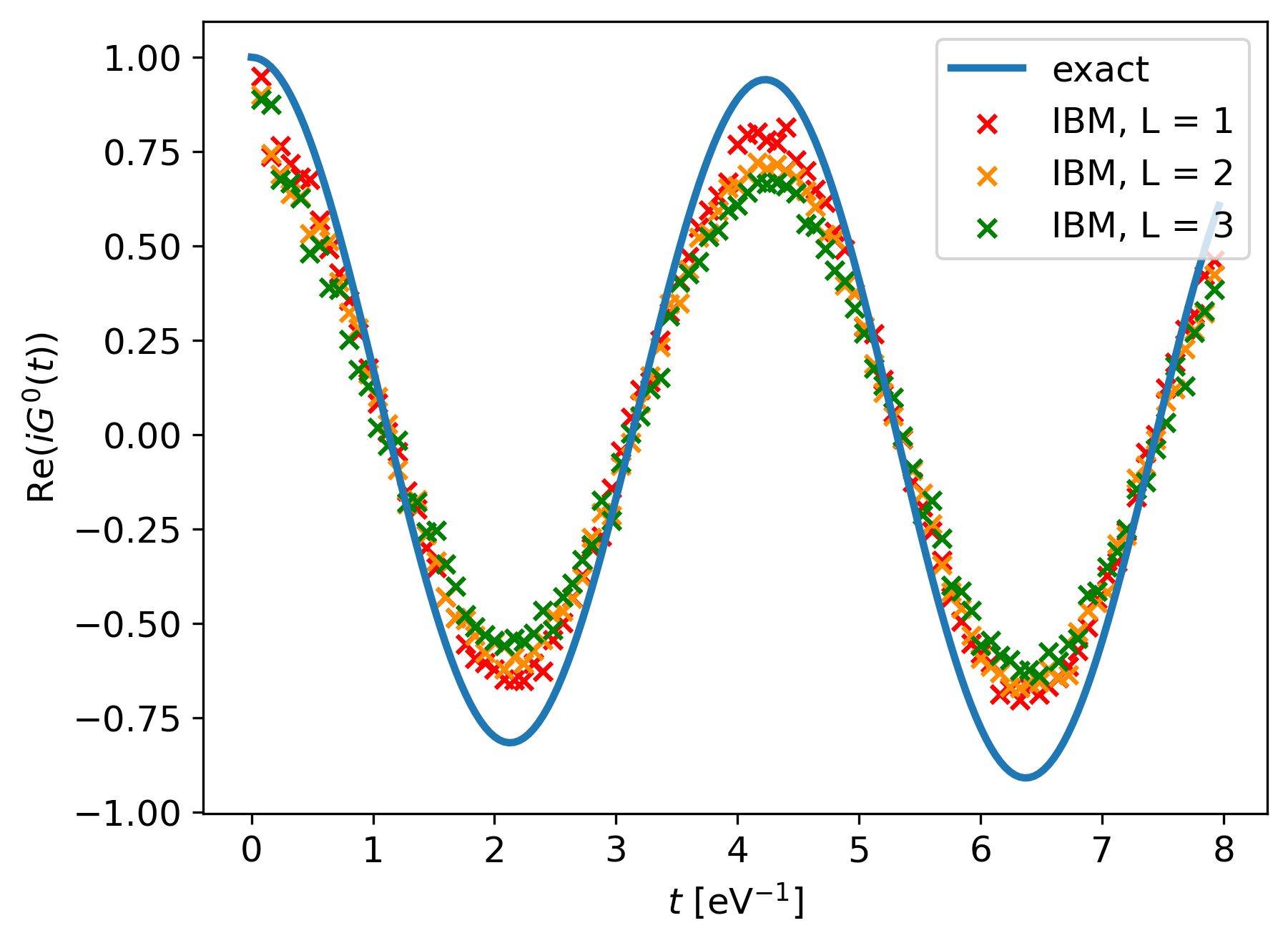}
    \caption{Real component of $G^{0}(t)$ evaluated on IBM Heron (ibm\_torino) for $L\in\{1,2,3\}$ compared to the exact analytic calculation showing the increase in damping dependent on $L$.}
    \label{fig:U0results}
\end{figure}
As stated above, we also compile circuits to measure $G^0(t)$ to verify the damping effect.
We used the same ansatzes so that the circuit depth and the effects caused by the damping are comparable. 
Compared to the actual circuits, they have slightly fewer $C_Z$-gates (61 instead of 65 for $L=3$) for later points caused by small parameter values for certain gates that are therefore filtered out by the compiler.
We found that the Green's function $G^0(t)$ for this Hamiltonian is correctly approximated by all three cases at least up to $t_\text{max}$ and that the damping varies only slightly over time, as the entanglement of the system is restricted to each spin sector.
We can therefore calculate the coefficients $\alpha_L$ and $\alpha_\text{GS}$ by comparing the measured results for $G^0_L(t)$ with the analytical expression, illustrated in Fig. \ref{fig:U0results}.
Calculating the damping factors around the peaks of $G^0(t)$ allows us to improve the results of $G(t)$ (see Fig. \ref{fig:hardwareResults}(b)) using the method described in Sec. \ref{subsec:errormitigation}. 

\begin{figure}[t]
    \centering
    \includegraphics[width = 1\linewidth]{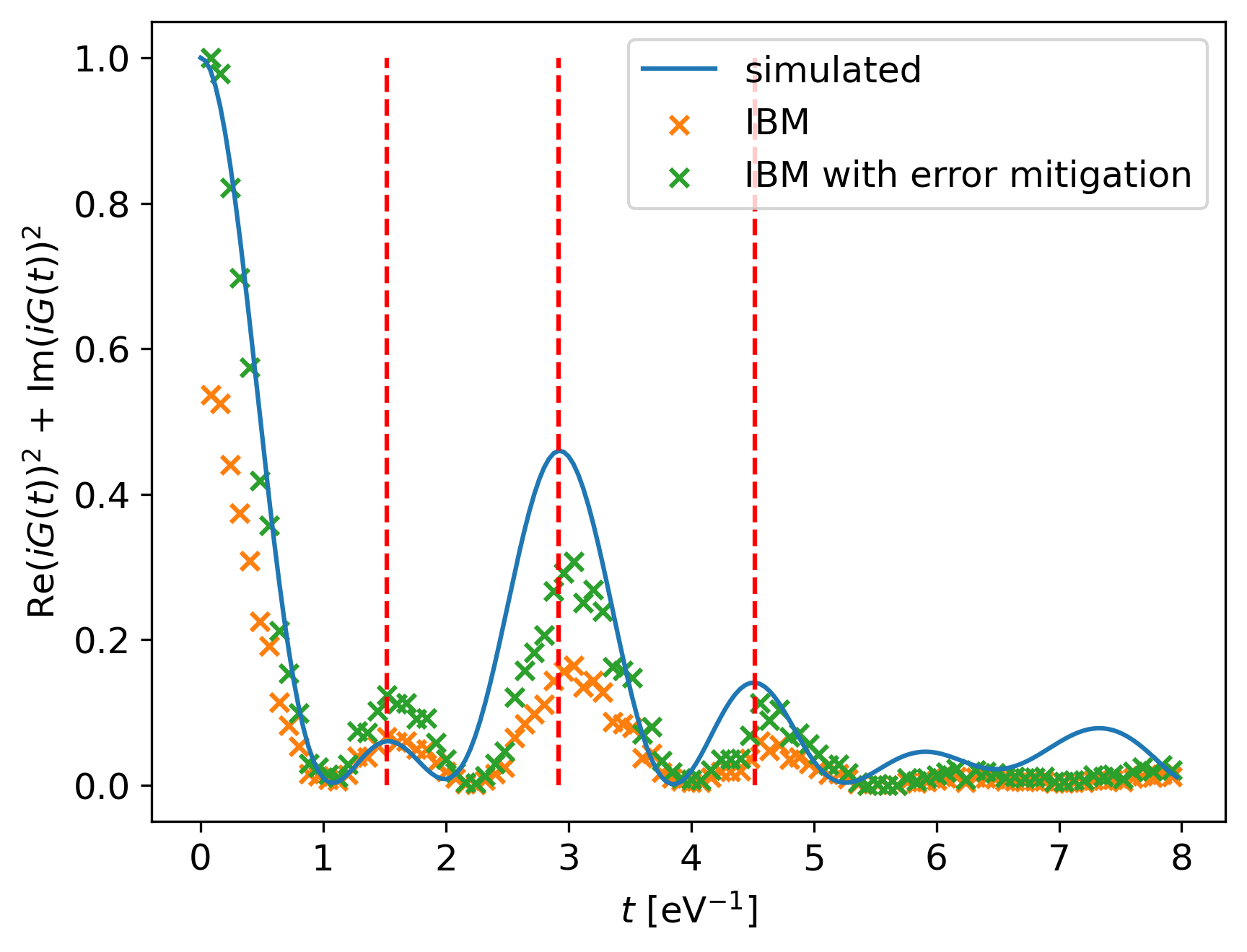}
    \caption{Envelope of $iG(t)$. The Green's function measured on the quantum device follows the expected behavior but exhibits a temporal delay and an amplitude discrepancy.}
    \label{envelop}
\end{figure}
Although the measured results follow the curve accurately for most points in time, they differ for certain time ranges.
We look at the Green's function envelope given by
\begin{equation}
    |iG(t)|^2 = \text{Re}(iG(t))^2 + \text{Im}(iG(t))^2
\end{equation}
to analyze the occurring errors. 
While the envelope decays to zero for a system with an infinite number of bath sites, finite-size systems such as ours show an initial decay followed by periodically appearing quantum revivals, called Poincaré recurrences (see Fig. \ref{envelop}).  
This is caused by the finite number of degrees of freedom, which results in a discrete energy spectrum. 

The hardware result captures the dynamics of these revivals, but exhibits two distinct deviations from the theoretical prediction: the expected damping effect visible in the amplitudes of the peaks of the envelope and also a horizontal delay illustrated by the shifts of the peaks in the envelope to the right.
As expected, our method of rescaling for handling the former error corrects the curve for the initial points; however, it either overcompensates or is not sufficient for other time ranges (see Fig. \ref{fig:hardwareResults}).
The latter error, the peak shift, is caused by the order in which we run the circuits.
Since the circuits for the real and imaginary components are executed sequentially on the hardware, this delay suggests the presence of device calibration drift during the execution window. 
Such drift typically alters the physical qubit frequencies, resulting in systematic under-rotation of the two-qubit entangling gates. 
This systematic error physically manifests itself as weakened effective coupling strengths within the simulated Hamiltonian, thereby retarding the propagation of correlations across the qubit chain and delaying the observed revivals.

\begin{figure*}[t]
    \centering
    \includegraphics[width=1\textwidth]{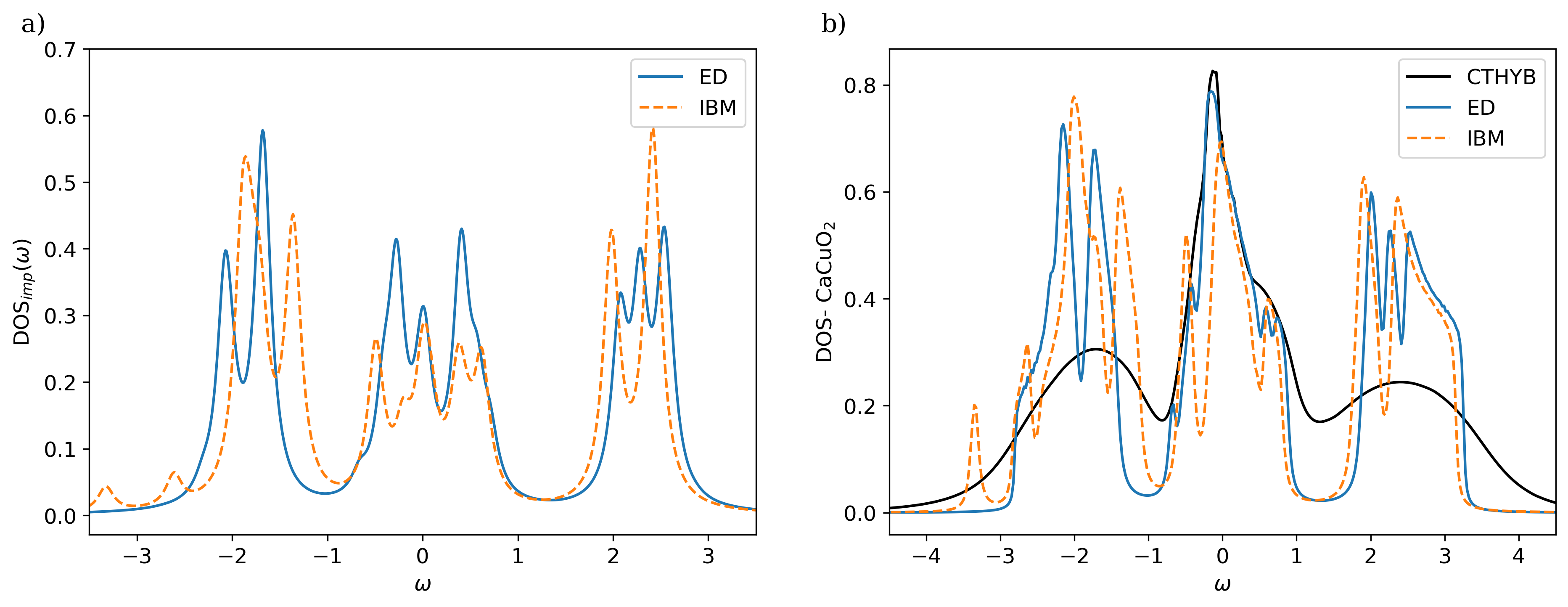}
    \caption{Density of states of the impurity model (a) and for \ce{CaCuO2} after embedding (b) with $\eta = 0.1$. We compare our hardware results (IBM) in both cases with the exact diagonalization (ED) and, for (b) additionally with Quantum Monte Carlo (CTHYB) results after analytical continuation to the real-frequency axis. Although the individual peaks cannot be resolved more accurately, the separation of the quasiparticle peak from the lower and upper Hubbard bands below and above the Fermi level, respectively, is clearly visible.}
    \label{dos}
\end{figure*}

Nevertheless, these are errors that we can only analyze because we have access to the simulated results. 
We restrict our further calculations without correcting these errors, as would be the case for an experiment with system sizes exceeding the capacities of classical hardware. 
With our postprocessing technique (see Appendix \ref{app:postprocessing}), which would allow for accurate results using state-vector simulations, we are able to find some $\omega_j$ such that we can reconstruct the density of states. 
As illustrated in Fig. \ref{dos}, the obtained DOS, even though the procedure is robust against noise, suffers from the aforementioned errors caused by the hardware. 
However, the split into three distinct peaks, the lower and upper Hubbard bands, and the quasiparticle peak, can be recovered.
To verify the quality of our result, we calculate the quasiparticle weight (QPW) as described in Sec.~\ref{subsec:qpw} and
 summarize our findings in Table \ref{tab:qpw}.
The QPW calculated from our noisy Green's function obtained from our hardware measurements differs by only 15 \% from the state-vector simulation.
\begin{table}[t]
\def\arraystretch{1.5}
\begin{tabular}{c|c}
    Method & QPW\\\hline
    Quantum Monte Carlo (QMC) & 0.4453\\\hline
    Exact Diagonalization (ED) & 0.3796\\
    Simulated with Compression ($L=3$, up to $t_\text{max})$  & 0.3560\\\hline
    IBM Heron & 0.3027
    \end{tabular}
    \caption{Quasiparticle weight calculated by integration over the central quasiparticle peak for the listed methods using the same number of bath sites. Even with just a short time window, the simulated results show good accuracy compared to exact diagonalization. Noise on the hardware reduces the value by 15 \% compared to the simulated result. }
\label{tab:qpw}
\end{table}

\section{Discussion}
\label{sec:discussion}
In this work, we presented a hybrid quantum-classical workflow for computing the electronic properties of \ce{CaCuO2}, a parent structure of high-temperature cuprate superconductors, utilizing a quantum impurity solver based on variational quantum compressed time evolution. 
By evaluating the time-dependent Green's function $G(t)$ on the IBM Heron quantum device (ibm\_torino), we successfully captured the system dynamics. 
Utilizing a Lehmann representation fitting procedure, we demonstrated that it is possible to reconstruct the density of states from these short-time hardware measurements. 
Despite the presence of hardware noise, the post-processed DOS successfully recovered the structure of the strongly correlated material: the lower and upper Hubbard bands, and the central quasiparticle peak.
However, an in-depth analysis of the measured Green's function envelope reveals significant limitations imposed by current NISQ-era hardware.

We observed a substantial damping effect at later simulation times, driven by dephasing errors that destroy information as entanglement within the system grows, which are not accurately captured by the hardware. 
Additionally, the experimental data exhibited a temporal delay relative to the exact solution. 
This issue can be circumvented for the evaluation of $G(t)$ by reordering the time points for the calculation on the quantum device, but it heavily restricts the ability to train the parametrized circuits directly on hardware.
During gradient-descent optimization, the circuit parameters change only slightly between iterations, just as the parameters of the circuits for consecutive time steps do. 
Gradient estimates would therefore suffer from the same hardware drift observed in this work.
Consequently, training on current hardware is infeasible for long time scales, where errors from the variational time evolution compression accumulate. 
While our error mitigation technique, based on extrapolating measurements of the non-interacting systems, succeeds in correcting early-time amplitudes, it either overcompensates or undercompensates for errors at later times and cannot correct frequency shifts. 

Hence, while our methodology is sufficiently robust to extract qualitative features of the spectral function from noisy data, the current quantum hardware does not allow for a complete dynamical mean-field theory calculation. 
A full DMFT procedure requires strict self-consistency, meaning the output Green's function must be iteratively fed back into the bath fitting equations. 
For the DMFT loop to converge reliably, much higher numerical accuracy is needed than what is currently achievable. 
This requires longer simulation times, which are currently not feasible on quantum devices, even when utilizing time evolution compression. 
To transition this proof-of-concept into a tool capable of executing the full, self-consistent DFT+DMFT workflow for real materials, including the training of the required parametrized circuits, hardware beyond the NISQ era will be required. 
Once quantum hardware reaches this level of maturity, quantum-device implementations of DMFT for larger systems and clusters could become competitive with classical methods such as DMRG, providing a useful tool for the research of strongly correlated materials.  

\begin{acknowledgments}
This work was supported by the German Federal Ministry of Research, Technology and Space (BMFTR) within the funding program “Quantum technologies — From basic research to market” in the projects MANIQU (Grant No. 13N15577) and EQUAHUMO (Grant No. 13N16067), and the Munich Quantum Valley, which is supported by the Bavarian state government with funds from the Hightech Agenda Bayern Plus. 
\end{acknowledgments}
\newpage
\bibliography{bib}

@article{Georges,
  title = {Dynamical mean-field theory of strongly correlated fermion systems and the limit of infinite dimensions},
  author = {Georges, Antoine and Kotliar, Gabriel and Krauth, Werner and Rozenberg, Marcelo J.},
  journal = {Rev. Mod. Phys.},
  volume = {68},
  issue = {1},
  pages = {13--125},
  numpages = {0},
  year = {1996},
  month = {Jan},
  publisher = {American Physical Society},
  doi = {10.1103/RevModPhys.68.13},
  url = {https://link.aps.org/doi/10.1103/RevModPhys.68.13}
}

@article{Ehrlich,
    doi = {10.1088/1361-648X/add4db},
    url = {https://doi.org/10.1088/1361-648X/add4db},
    year = {2025},
    month = {may},
    publisher = {IOP Publishing},
    volume = {37},
    number = {22},
    pages = {225901},
    author = {Ehrlich, Jannis and Urban, Daniel F and Elsässer, Christian},
    title = {Variational quantum-algorithm based self-consistent calculations for the two-site {DMFT} model on noisy quantum computing hardware},
    journal = {J. Phys.: Condens. Matter}
}

@article{Endo,
  title = {Calculation of the {Green's} function on near-term quantum computers},
  author = {Endo, Suguru and Kurata, Iori and Nakagawa, Yuya O.},
  journal = {Phys. Rev. Res.},
  volume = {2},
  issue = {3},
  pages = {033281},
  numpages = {17},
  year = {2020},
  month = {Aug},
  publisher = {American Physical Society},
  doi = {10.1103/PhysRevResearch.2.033281},
  url = {https://link.aps.org/doi/10.1103/PhysRevResearch.2.033281}
}

@misc{Jamet,
      title={Quantum subspace expansion algorithm for {Green's} functions}, 
      author={Francois Jamet and Abhishek Agarwal and Ivan Rungger},
      year={2022},
      eprint={2205.00094},
      archivePrefix={arXiv},
      primaryClass={quant-ph},
      url={https://arxiv.org/abs/2205.00094}, 
}

@article{Rizzo,
  title = {One-particle {Green's} functions from the quantum equation of motion algorithm},
  author = {Rizzo, Jacopo and Libbi, Francesco and Tacchino, Francesco and Ollitrault, Pauline J. and Marzari, Nicola and Tavernelli, Ivano},
  journal = {Phys. Rev. Res.},
  volume = {4},
  issue = {4},
  pages = {043011},
  numpages = {10},
  year = {2022},
  month = {Oct},
  publisher = {American Physical Society},
  doi = {10.1103/PhysRevResearch.4.043011},
  url = {https://link.aps.org/doi/10.1103/PhysRevResearch.4.043011}
}

@misc{Rungger,
      title={Dynamical mean field theory algorithm and experiment on quantum computers}, 
      author={I. Rungger and N. Fitzpatrick and H. Chen and C. H. Alderete and H. Apel and A. Cowtan and A. Patterson and D. Munoz Ramo and Y. Zhu and N. H. Nguyen and E. Grant and S. Chretien and L. Wossnig and N. M. Linke and R. Duncan},
      year={2020},
      eprint={1910.04735},
      archivePrefix={arXiv},
      primaryClass={quant-ph},
      url={https://arxiv.org/abs/1910.04735}, 
}

@article{Selisko,
  author    = {Selisko, Johannes and Amsler, Maximilian and Wever, Christopher and Kawashima, Yukio and Samsonidze, Georgy and Ul Haq, Rukhsan and Tacchino, Francesco and Tavernelli, Ivano and Eckl, Thomas},
  title     = {Dynamical mean field theory for real materials on a quantum computer},
  journal   = {npj Comput. Mater.},
  volume    = {11},
  number    = {1},
  pages     = {325},
  year      = {2025},
  doi       = {10.1038/s41524-025-01772-6},
  url       = {https://www.nature.com/articles/s41524-025-01772-6}
}

@article{Bauer,
  title = {Hybrid Quantum-Classical Approach to Correlated Materials},
  author = {Bauer, Bela and Wecker, Dave and Millis, Andrew J. and Hastings, Matthew B. and Troyer, Matthias},
  journal = {Phys. Rev. X},
  volume = {6},
  issue = {3},
  pages = {031045},
  numpages = {11},
  year = {2016},
  month = {Sep},
  publisher = {American Physical Society},
  doi = {10.1103/PhysRevX.6.031045},
  url = {https://link.aps.org/doi/10.1103/PhysRevX.6.031045}
}

@article{Hogan,
  title     = {Efficient quantum implementation of dynamical mean field theory for correlated materials},
  author    = {Hogan, Norman and K{\"{o}}kc{\"{u}}, Efekan and Steckmann, Thomas and Doak, Liam P. and Mejuto-Zaera, Carlos and Camps, Daan and Van Beeumen, Roel and de Jong, Wibe A. and Kemper, A. F.},
  journal   = {npj Comput. Mater.},
  volume    = {12},
  year      = {2026},
  doi       = {10.1038/s41524-026-02289-2},
  eprint    = {2508.05738},
  archivePrefix = {arXiv},
  primaryClass = {quant-ph}
}

@article{Jaderberg,
doi = {10.1088/2058-9565/ab972b},
url = {https://dx.doi.org/10.1088/2058-9565/ab972b},
year = {2020},
month = {jun},
publisher = {IOP Publishing},
volume = {5},
number = {3},
pages = {034015},
author = {B Jaderberg and A Agarwal and K Leonhardt and M Kiffner and D Jaksch},
title = {Minimum hardware requirements for hybrid quantum–classical {DMFT}},
journal = {Quantum Sci. Technol.},
}

@article{Keen,
    doi = {10.1088/2058-9565/ab7d4c},
    url = {https://dx.doi.org/10.1088/2058-9565/ab7d4c},
    year = {2020},
    month = {apr},
    publisher = {IOP Publishing},
    volume = {5},
    number = {3},
    pages = {035001},
    author = {Trevor Keen and Thomas Maier and Steven Johnston and Pavel Lougovski},
    title = {Quantum-classical simulation of two-site dynamical mean-field theory on noisy quantum hardware},
    journal = {Quantum Sci. Technol.}
}

@article{Kreula,
  author    = {Kreula, J. M. and Garc{\'i}a-\'Alvarez, L. and Lamata, L. and Clark, S. R. and Solano, E. and Jaksch, D.},
  title     = {Few-qubit quantum-classical simulation of strongly correlated lattice fermions},
  journal   = {EPJ Quantum Technol.},
  year      = {2016},
  volume    = {3},
  number    = {1},
  pages     = {11},
  doi       = {10.1140/epjqt/s40507-016-0049-1},
  url       = {https://doi.org}
}

@article{Steckmann,
  title = {Mapping the metal-insulator phase diagram by algebraically fast-forwarding dynamics on a cloud quantum computer},
  author = {Steckmann, Thomas and Keen, Trevor and K\"okc\"u, Efekan and Kemper, Alexander F. and Dumitrescu, Eugene F. and Wang, Yan},
  journal = {Phys. Rev. Res.},
  volume = {5},
  issue = {2},
  pages = {023198},
  numpages = {16},
  year = {2023},
  month = {Jun},
  publisher = {American Physical Society},
  doi = {10.1103/PhysRevResearch.5.023198},
  url = {https://link.aps.org/doi/10.1103/PhysRevResearch.5.023198}
}

@misc{Wolf,
      title={Variational Time Evolution Compression for Solving Impurity Models on Quantum Hardware}, 
      author={Stefan Wolf and Martin Eckstein and Michael J. Hartmann},
      year={2025},
      eprint={2508.10526},
      archivePrefix={arXiv},
      primaryClass={quant-ph},
      url={https://arxiv.org/abs/2508.10526}, 
}

@article{Barison,
  doi = {10.22331/q-2021-07-28-512},
  url = {https://doi.org/10.22331/q-2021-07-28-512},
  title = {An efficient quantum algorithm for the time evolution of parameterized circuits},
  author = {Barison, Stefano and Vicentini, Filippo and Carleo, Giuseppe},
  journal = {{Quantum}},
  issn = {2521-327X},
  publisher = {{Verein zur F{\"{o}}rderung des Open Access Publizierens in den Quantenwissenschaften}},
  volume = {5},
  pages = {512},
  month = jul,
  year = {2021}
}

@article{Benedetti,
  title = {Hardware-efficient variational quantum algorithms for time evolution},
  author = {Benedetti, Marcello and Fiorentini, Mattia and Lubasch, Michael},
  journal = {Phys. Rev. Res.},
  volume = {3},
  issue = {3},
  pages = {033083},
  numpages = {15},
  year = {2021},
  month = {Jul},
  publisher = {American Physical Society},
  doi = {10.1103/PhysRevResearch.3.033083},
  url = {https://link.aps.org/doi/10.1103/PhysRevResearch.3.033083}
}

@article{Berthusen,
  title = {Quantum dynamics simulations beyond the coherence time on noisy intermediate-scale quantum hardware by variational {Trotter} compression},
  author = {Berthusen, Noah F. and Trevisan, Tha\'{\i}s V. and Iadecola, Thomas and Orth, Peter P.},
  journal = {Phys. Rev. Res.},
  volume = {4},
  issue = {2},
  pages = {023097},
  numpages = {12},
  year = {2022},
  month = {May},
  publisher = {American Physical Society},
  doi = {10.1103/PhysRevResearch.4.023097},
  url = {https://link.aps.org/doi/10.1103/PhysRevResearch.4.023097}
}

@article{Goh,
  title = {Direct estimation of the density of states for fermionic systems},
  author = {Goh, Matthew L. and Koczor, B\'alint},
  journal = {Phys. Rev. Res.},
  volume = {8},
  issue = {1},
  pages = {013250},
  numpages = {18},
  year = {2026},
  month = {Mar},
  publisher = {American Physical Society},
  doi = {10.1103/h8l5-87zt},
  url = {https://link.aps.org/doi/10.1103/h8l5-87zt}
}

@article{Puig,
  title = {Variational Quantum Simulation: A Case Study for Understanding Warm Starts},
  author = {Puig, Ricard and Drudis, Marc and Thanasilp, Supanut and Holmes, Zo\"e},
  journal = {PRX Quantum},
  volume = {6},
  issue = {1},
  pages = {010317},
  numpages = {39},
  year = {2025},
  month = {Jan},
  publisher = {American Physical Society},
  doi = {10.1103/PRXQuantum.6.010317},
  url = {https://link.aps.org/doi/10.1103/PRXQuantum.6.010317}
}

@article{Bauernfeind2017,
  title = {Fork Tensor-Product States: Efficient Multiorbital Real-Time {DMFT} Solver},
  author = {Bauernfeind, Daniel and Zingl, Manuel and Triebl, Robert and Aichhorn, Markus and Evertz, Hans Gerd},
  journal = {Phys. Rev. X},
  volume = {7},
  issue = {3},
  pages = {031013},
  numpages = {15},
  year = {2017},
  month = {Jul},
  publisher = {American Physical Society},
  doi = {10.1103/PhysRevX.7.031013},
  url = {https://link.aps.org/doi/10.1103/PhysRevX.7.031013}
}

@article{Bauernfeind2018,
  title = {Dynamical mean-field theory on the real-frequency axis: $p\text{\ensuremath{-}}d$ hybridization and atomic physics in {${\mathrm{SrMnO}}_{3}$}},
  author = {Bauernfeind, Daniel and Triebl, Robert and Zingl, Manuel and Aichhorn, Markus and Evertz, Hans Gerd},
  journal = {Phys. Rev. B},
  volume = {97},
  issue = {11},
  pages = {115156},
  numpages = {11},
  year = {2018},
  month = {Mar},
  publisher = {American Physical Society},
  doi = {10.1103/PhysRevB.97.115156},
  url = {https://link.aps.org/doi/10.1103/PhysRevB.97.115156}
}

@article{Kugler,
  title = {Strongly Correlated Materials from a Numerical Renormalization Group Perspective: How the {Fermi-liquid} State of {${\mathrm{Sr}}_{2}{\mathrm{RuO}}_{4}$} Emerges},
  author = {Kugler, Fabian B. and Zingl, Manuel and Strand, Hugo U. R. and Lee, Seung-Sup B. and von Delft, Jan and Georges, Antoine},
  journal = {Phys. Rev. Lett.},
  volume = {124},
  issue = {1},
  pages = {016401},
  numpages = {7},
  year = {2020},
  month = {Jan},
  publisher = {American Physical Society},
  doi = {10.1103/PhysRevLett.124.016401},
  url = {https://link.aps.org/doi/10.1103/PhysRevLett.124.016401}
}

@article{Stadler,
  title = {Dynamical Mean-Field Theory Plus Numerical Renormalization-Group Study of Spin-Orbital Separation in a Three-Band {Hund} Metal},
  author = {Stadler, K. M. and Yin, Z. P. and von Delft, J. and Kotliar, G. and Weichselbaum, A.},
  journal = {Phys. Rev. Lett.},
  volume = {115},
  issue = {13},
  pages = {136401},
  numpages = {5},
  year = {2015},
  month = {Sep},
  publisher = {American Physical Society},
  doi = {10.1103/PhysRevLett.115.136401},
  url = {https://link.aps.org/doi/10.1103/PhysRevLett.115.136401}
}

@article{WolfCheby,
  title = {Chebyshev matrix product state impurity solver for dynamical mean-field theory},
  author = {Wolf, F. Alexander and McCulloch, Ian P. and Parcollet, Olivier and Schollw\"ock, Ulrich},
  journal = {Phys. Rev. B},
  volume = {90},
  issue = {11},
  pages = {115124},
  numpages = {17},
  year = {2014},
  month = {Sep},
  publisher = {American Physical Society},
  doi = {10.1103/PhysRevB.90.115124},
  url = {https://link.aps.org/doi/10.1103/PhysRevB.90.115124}
}

@article{Arovas,
   title={The {Hubbard} Model},
   volume={13},
   ISSN={1947-5462},
   url={http://dx.doi.org/10.1146/annurev-conmatphys-031620-102024},
   DOI={10.1146/annurev-conmatphys-031620-102024},
   number={1},
   journal={Annu. Rev. Condens. Matter Phys.},
   publisher={Annual Reviews},
   author={Arovas, Daniel P. and Berg, Erez and Kivelson, Steven A. and Raghu, Srinivas},
   year={2022},
   month=Mar, pages={239–274} }

@article{Gull,
  title = {Continuous-time {Monte Carlo} methods for quantum impurity models},
  author = {Gull, Emanuel and Millis, Andrew J. and Lichtenstein, Alexander I. and Rubtsov, Alexey N. and Troyer, Matthias and Werner, Philipp},
  journal = {Rev. Mod. Phys.},
  volume = {83},
  issue = {2},
  pages = {349--404},
  numpages = {0},
  year = {2011},
  month = {May},
  publisher = {American Physical Society},
  doi = {10.1103/RevModPhys.83.349},
  url = {https://link.aps.org/doi/10.1103/RevModPhys.83.349}
}

@article{giannozzi_quantum_2009,
	title = {{QUANTUM} {ESPRESSO}: a modular and open-source software project for quantum simulations of materials},
	volume = {21},
	issn = {0953-8984},
	shorttitle = {{QUANTUM} {ESPRESSO}},
	url = {https://dx.doi.org/10.1088/0953-8984/21/39/395502},
	doi = {10.1088/0953-8984/21/39/395502},
	number = {39},
	urldate = {2022-10-28},
	journal = {J. Phys.: Condens. Matter},
	author = {Giannozzi, Paolo and Baroni, Stefano and Bonini, Nicola and Calandra, Matteo and Car, Roberto and Cavazzoni, Carlo and Ceresoli, Davide and Chiarotti, Guido L. and Cococcioni, Matteo and Dabo, Ismaila and {Dal Corso}, Andrea and {de Gironcoli}, Stefano and Fabris, Stefano and Fratesi, Guido and Gebauer, Ralph and Gerstmann, Uwe and Gougoussis, Christos and Kokalj, Anton and Lazzeri, Michele and Martin-Samos, Layla and Marzari, Nicola and Mauri, Francesco and Mazzarello, Riccardo and Paolini, Stefano and Pasquarello, Alfredo and Paulatto, Lorenzo and Sbraccia, Carlo and Scandolo, Sandro and Sclauzero, Gabriele and Seitsonen, Ari P. and Smogunov, Alexander and Umari, Paolo and Wentzcovitch, Renata M.},
	month = sep,
	year = {2009},
	pages = {395502},
}

@article{Jordan,
  author  = {Jordan, P. and Wigner, E.},
  title   = {{Ü}ber das {P}aulische {Ä}quivalenzverbot},
  journal = {Z. Phys.},
  volume  = {47},
  number  = {9},
  pages   = {631--651},
  year    = {1928},
  doi     = {10.1007/BF01331938}
}

@article{Keimer,
    author = {Keimer, B. and Kivelson, S. A. and Norman, M. R. and Uchida, S. and Zaanen, J.},
    title = {From quantum matter to high-temperature superconductivity in copper oxides},
    volume = {518},
    url = {https://doi.org/10.1038/nature14165},
    doi = {10.1038/nature14165},
    pages = {179--186},
    year = {2015},
    issn = {7538},
    journal = {Nature},
    month = feb
}

@article{Kotliar,
  title = {Electronic structure calculations with dynamical mean-field theory},
  author = {Kotliar, G. and Savrasov, S. Y. and Haule, K. and Oudovenko, V. S. and Parcollet, O. and Marianetti, C. A.},
  journal = {Rev. Mod. Phys.},
  volume = {78},
  issue = {3},
  pages = {865--951},
  numpages = {0},
  year = {2006},
  month = {Aug},
  publisher = {American Physical Society},
  doi = {10.1103/RevModPhys.78.865},
  url = {https://link.aps.org/doi/10.1103/RevModPhys.78.865}
}

@article{Norman,
   title={The Challenge of Unconventional Superconductivity},
   volume={332},
   ISSN={1095-9203},
   url={http://dx.doi.org/10.1126/science.1200181},
   DOI={10.1126/science.1200181},
   number={6026},
   journal={Science},
   publisher={American Association for the Advancement of Science (AAAS)},
   author={Norman, Michael R.},
   year={2011},
   month=Apr, pages={196–200} }

@article{perdew_generalized_1996,
 author = {Perdew, John P. and Burke, Kieron and Ernzerhof, Matthias},
 title = {Generalized Gradient Approximation Made Simple},
 volume = {77},
 url = {https://doi.org/10.1103/physrevlett.77.3865},
 doi = {10.1103/physrevlett.77.3865},
 number = {18},
 urldate = {2022-10-28},
 journal = {Phys. Rev. Lett.},
 month = oct,
 year = {1996},
 pages = {3865--3868},
 source = {Crossref},
 publisher = {American Physical Society (APS)},
 issn = {0031-9007, 1079-7114},
}

@article{Peruzzo,
  author    = {Peruzzo, Alberto and McClean, Jarrod and Shadbolt, Peter and Yung, Man-Hong and Zhou, Xiao-Qi and Love, Peter J. and Aspuru-Guzik, Al{\'a}n and O'Brien, Jeremy L.},
  title     = {A variational eigenvalue solver on a photonic quantum processor},
  journal   = {Nat. Commun.},
  volume    = {5},
  number    = {1},
  pages     = {4213},
  year      = {2014},
  doi       = {10.1038/ncomms5213},
  url       = {https://www.nature.com/articles/ncomms5213}
}

@article{van_setten_pseudodojo_2018,
 author = {van Setten, M.J. and Giantomassi, M. and Bousquet, E. and Verstraete, M.J. and Hamann, D.R. and Gonze, X. and Rignanese, G.-M.},
 title = {The {PseudoDojo:} {Training} and grading a 85 element optimized norm-conserving pseudopotential table},
 volume = {226},
 issn = {0010-4655},
 shorttitle = {The PseudoDojo},
 url = {https://doi.org/10.1016/j.cpc.2018.01.012},
 doi = {10.1016/j.cpc.2018.01.012},
 urldate = {2022-10-28},
 journal = {Comput. Phys. Commun.},
 month = may,
 year = {2018},
 pages = {39--54},
 source = {Crossref},
 publisher = {Elsevier BV},
}

@article{pizzi_wannier90_2020,
	title = {Wannier90 as a community code: new features and applications},
	volume = {32},
	issn = {0953-8984},
	shorttitle = {Wannier90 as a community code},
	url = {https://dx.doi.org/10.1088/1361-648X/ab51ff},
	doi = {10.1088/1361-648X/ab51ff},
	number = {16},
	urldate = {2022-10-28},
	journal = {J. Phys.: Condens. Matter},
	author = {Pizzi, Giovanni and Vitale, Valerio and Arita, Ryotaro and  Blügel, Stefan and Freimuth, Frank and Géranton, Guillaume and Gibertini, Marco and Gresch, Dominik and Johnson, Charles and Koretsune, Takashi and Ibañez-Azpiroz, Julen and Lee, Hyungjun and Lihm, Jae-Mo and Marchand, Daniel and Marrazzo, Antimo and Mokrousov, Yuriy and Mustafa, Jamal I. and Nohara, Yoshiro and Nomura, Yusuke and Paulatto, Lorenzo and Poncé, Samuel and Ponweiser, Thomas and Qiao, Junfeng and Thöle, Florian and Tsirkin, Stepan S. and Wierzbowska, MaÅgorzata and Marzari, Nicola and Vanderbilt, David and Souza, Ivo and Mostofi, Arash A. and Yates, Jonathan R.},
	month = jan,
	year = {2020},
	pages = {165902},
}

@article{nakamura-20-respack,
 author = {Nakamura, Kazuma and Yoshimoto, Yoshihide and Nomura, Yusuke and Tadano, Terumasa and Kawamura, Mitsuaki and Kosugi, Taichi and Yoshimi, Kazuyoshi and Misawa, Takahiro and Motoyama, Yuichi},
 title = {{RESPACK:} {An} ab initio tool for derivation of effective low-energy model of material},
 journal = {Comput. Phys. Commun.},
 volume = {261},
 pages = {107781},
 year = {2021},
 issn = {0010-4655},
 doi = {10.1016/j.cpc.2020.107781},
 url = {https://doi.org/10.1016/j.cpc.2020.107781},
 source = {Crossref},
 publisher = {Elsevier BV},
 month = apr,
}

@article{kurita_interface_2023,
	title = {Interface tool from {Wannier90} to {RESPACK}: wan2respack},
	volume = {292},
	issn = {0010-4655},
	shorttitle = {Interface tool from {Wannier90} to {RESPACK}},
	url = {https://www.sciencedirect.com/science/article/pii/S0010465523001996},
	doi = {10.1016/j.cpc.2023.108854},
	urldate = {2024-03-08},
	journal = {Comput. Phys. Commun.},
	author = {Kurita, Kensuke and Misawa, Takahiro and Yoshimi, Kazuyoshi and Ido, Kota and Koretsune, Takashi},
	month = nov,
	year = {2023},
	pages = {108854},
}

@article{Merkel2022, 
    doi = {10.21105/joss.04623}, 
    url = {https://doi.org/10.21105/joss.04623}, 
    year = {2022}, 
    publisher = {The Open Journal}, 
    volume = {7}, 
    number = {77}, 
    pages = {4623}, 
    author = {Maximilian E. Merkel and Alberto Carta and Sophie Beck and Alexander Hampel}, 
    title = {solid\_dmft: gray-boxing {DFT+DMFT} materials simulations with {TRIQS}}, 
    journal = {J. Open Source Softw.} 
}

@article{beck_charge_2022,
	title = {Charge self-consistent electronic structure calculations with dynamical mean-field theory using {Quantum} {ESPRESSO}, {Wannier}90 and {TRIQS}},
	volume = {34},
	issn = {0953-8984},
	url = {https://dx.doi.org/10.1088/1361-648X/ac5d1c},
	doi = {10.1088/1361-648X/ac5d1c},
	number = {23},
	urldate = {2024-03-04},
	journal = {J. Phys.: Condens. Matter},
	author = {Beck, Sophie and Hampel, Alexander and Parcollet, Olivier and Ederer, Claude and Georges, Antoine},
	month = apr,
	year = {2022},
	pages = {235601},
}

@article{seth_triqscthyb_2016,
	title = {{TRIQS}/{CTHYB}: {A} continuous-time quantum {Monte} {Carlo} hybridisation expansion solver for quantum impurity problems},
	volume = {200},
	issn = {0010-4655},
	shorttitle = {{TRIQS}/{CTHYB}},
	url = {https://www.sciencedirect.com/science/article/pii/S001046551500404X},
	doi = {10.1016/j.cpc.2015.10.023},
	urldate = {2024-03-01},
	journal = {Comput. Phys. Commun.},
	author = {Seth, Priyanka and Krivenko, Igor and Ferrero, Michel and Parcollet, Olivier},
	month = mar,
	year = {2016},
	pages = {274--284},
}

@article{triqs,
title = "{TRIQS}: A toolbox for research on interacting quantum systems",
journal = "Comput. Phys. Commun.",
volume = "196",
number = "",
pages = "398 - 415",
year = "2015",
note = "",
issn = "0010-4655",
doi = "http://dx.doi.org/10.1016/j.cpc.2015.04.023",
url = "http://www.sciencedirect.com/science/article/pii/S0010465515001666",
author = "Olivier Parcollet and Michel Ferrero and Thomas Ayral and Hartmut Hafermann and Igor Krivenko and Laura Messio and Priyanka Seth",
}

@article{Karp_CaCuO2,
  title = {Many-Body Electronic Structure of {${\mathrm{NdNiO}}_{2}$} and {${\mathrm{CaCuO}}_{2}$}},
  author = {Karp, Jonathan and Botana, Antia S. and Norman, Michael R. and Park, Hyowon and Zingl, Manuel and Millis, Andrew},
  journal = {Phys. Rev. X},
  volume = {10},
  issue = {2},
  pages = {021061},
  numpages = {11},
  year = {2020},
  month = {Jun},
  publisher = {American Physical Society},
  doi = {10.1103/PhysRevX.10.021061},
  url = {https://link.aps.org/doi/10.1103/PhysRevX.10.021061}
}

@article{Karp_CaCuO2_SC,
  title = {Superconductivity and antiferromagnetism in {${\mathrm{NdNiO}}_{2}$} and {${\mathrm{CaCuO}}_{2}$}: A cluster {DMFT} study},
  author = {Karp, Jonathan and Hampel, Alexander and Millis, Andrew J.},
  journal = {Phys. Rev. B},
  volume = {105},
  issue = {20},
  pages = {205131},
  numpages = {11},
  year = {2022},
  month = {May},
  publisher = {American Physical Society},
  doi = {10.1103/PhysRevB.105.205131},
  url = {https://link.aps.org/doi/10.1103/PhysRevB.105.205131}
}

@article{Lu_Efficient,
  title = {Efficient real-frequency solver for dynamical mean-field theory},
  author = {Lu, Y. and Höppner, M. and Gunnarsson, O. and Haverkort, M. W.},
  journal = {Phys. Rev. B},
  volume = {90},
  issue = {8},
  pages = {085102},
  numpages = {18},
  year = {2014},
  month = {Aug},
  publisher = {American Physical Society},
  doi = {10.1103/PhysRevB.90.085102},
  url = {https://link.aps.org/doi/10.1103/PhysRevB.90.085102}
}
\addcontentsline{toc}{section}{References}

\clearpage
\onecolumngrid
\appendix
\pagenumbering{alph}

\begin{appendix}

\section{Variational Time Evolution Compression}
\label{app:vqtec}
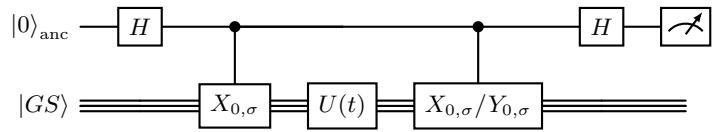
\begin{figure}[h]
\centering
\begin{quantikz}[wire types = {q, b}, classical gap=0.07cm]
    \lstick{$\ket{0}_\text{anc}$} & \gate{H} & \ctrl{1} & \qw  & \ctrl{1} & \gate{H} & \meter{} \\
    \lstick{$\ket{GS}$} & \qw & \gate{X_{0,\sigma}} & \gate{U(t)} & \gate{X_{0,\sigma}/Y_{0,\sigma}} & \qw & \qw 
\end{quantikz}
\caption{Hadamard test for measuring the Green's function w.r.t. the ground state on a quantum device}
\label{fig:hademardtest}
\end{figure}

\begin{figure}[h]
\centering
\begin{quantikz}[wire types = {q, b}, classical gap = 0.07cm]
    \lstick{$\ket{0}_\text{anc}$} & \gate{H} & \ctrl{1} & \qw  & \ctrl{1} & \gate{H} & \meter{} \\
    \lstick{$\ket{\text{GS}}$} & \qw & \gate{X_{0,\sigma}} & \gate{L_n} & \gate{X_{0,\sigma}} & \gate{U^\dag_\text{GS}} & \meter{}
\end{quantikz}
\caption{Circuit for measuring the cost function $C(\vec{\theta}_n)$ (Eq. \ref{eq:costfunction}).}
\label{fig:costmeasure}
\end{figure}
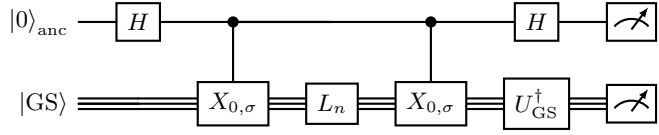

The single-impurity Anderson model in Eq. \ref{eq:SIAM} is formulated using fermionic annihilation and creation operators. 
For an application on a quantum device, it is necessary to transform these into quantum gates such that they still fulfill the anticommutation relations.
The most intuitive way to do this without increasing the number of qubits is the Jordan--Wigner transformation \cite{Jordan}:
\begin{align}
    c^\dag_{p,\sigma} \mapsto\frac{1}{2}Z_{0,\sigma}Z_{1,\sigma}\dots Z_{p-1,\sigma}\left(X_{p,\sigma}-iY_{p,\sigma}\right),\\
    c_{p,\sigma} \mapsto\frac{1}{2}Z_{0,\sigma}Z_{1,\sigma}\dots Z_{p-1,\sigma}\left(X_{p,\sigma}+iY_{p,\sigma}\right),
\end{align}
with
\begin{equation}
    X = \left(\begin{array}{cc}
         0 & 1 \\
         1 & 0
    \end{array}\right), \quad 
    Y = \left(\begin{array}{cc}
         0 & -i \\
         i & 0
    \end{array}\right), \quad
    Z = \left(\begin{array}{cc}
         1 & 0 \\
         0 & -1
    \end{array}\right).
\end{equation}
In this transformation, each mode is represented by one qubit, but it requires the $Z$ strings to keep track of the parity. 
Since the two spin sectors are coupled only via the density-density term $c^\dag_{0,\sigma}c_{0,\sigma}c^\dag_{0,\bar{\sigma}}c_{0,\bar{\sigma}}$, it is possible to perform this transformation for each spin sector separately.
The single-impurity Anderson model in chain topology then reads
\begin{align}
\begin{split}
    H &= 0.7775(\mathds{1}-Z_{0,\uparrow})(\mathds{1}-Z_{0,\downarrow}) - \sum_\sigma \left(0.655 (\mathds{1}-Z_{0,\sigma}) + 0.04(\mathds{1}-Z_{1,\sigma}) - 0.005 (\mathds{1}-Z_{2,\sigma})\right)\\
    & +\sum_\sigma\left[0.255(X_{0,\sigma}X_{1,\sigma} + Y_{0,\sigma}Y_{1,\sigma}) + 0.08 (X_{1,\sigma}X_{2,\sigma} + Y_{1,\sigma}Y_{2,\sigma})\right]
\end{split}
\end{align}
The time evolution operator can be approximated via a first-order Trotter expansion:
\begin{equation}
    U(t) = U(\Delta t)^N + \mathcal{O}(t²)
\end{equation}
with $\Delta t = \frac{t}{N}$ and
\begin{equation}
\begin{split}
    U(\Delta t) &= \left(\prod_\sigma R_{(XX+YY)_{01,\sigma}}(\Delta t \cdot 0.255)\right)\left(\prod_\sigma R_{(XX+YY)_{12,\sigma}}(\Delta t \cdot 0.08)\right)\,\\
    &\times\left(\prod_\sigma R_{Z_{2,\sigma}}(-\Delta t \cdot 0.005)R_{Z_{1,\sigma}}(\Delta t \cdot 0.04)R_{Z_{0,\sigma}}(-\Delta t \cdot 0.655)\right) C_P(\Delta t \cdot0.7775)
    \label{eq:singleTrotter}
\end{split}
\end{equation}
as illustrated in Fig. \ref{fig:HVA}. 
The gates are given by
\begin{equation*}
    C_P(\theta) = \left(\begin{array}{cccc}
         1&0&0&0  \\
         0&1&0&0 \\
         0&0&1&0 \\
         0&0&0&e^{i\theta}
    \end{array}\right)\quad \text{and} \quad 
    R_{XX+YY}(\theta) = \left(\begin{array}{cccc}
         1&0&0&0  \\
         0&\cos(\theta)&-i\sin(\theta)&0\\
         0&-i\sin(\theta)&\cos(\theta)&0\\
         0&0&0&1
    \end{array}\right).
\end{equation*}
The Green's function Eq. \ref{eq:timeGFJWT} can be measured using the Hadamard test as seen in Fig. \ref{fig:hademardtest}, which requires a single additional ancilla qubit.
Measuring $\braket{Z}_\text{anc}$ gives the real part of the expectation values required to calculate $G(t)$.
We use the same gate structure as in Eq. \ref{eq:singleTrotter} for our Hamiltonian variational ansatz $V(\vec{\theta})$ and replace the SIAM parameters with $\vec{\theta}$.
To evaluate the cost function Eq. \ref{eq:costfunction} to optimize the ansatz, we use a circuit similar to the Hadamard test (compare Fig. \ref{fig:costmeasure}).
We replace the time evolution operator with the training circuit $L_n$ and add an application of the inverse preparation of the ground state $U^\dag_\text{GS}$.
The state before the measurement reads
\begin{equation}
    \ket{\psi(\vec{\theta}_n)} = \frac{1}{2}\left(\ket{0}_\text{anc}\otimes U^\dag_\text{GS}(L_n + K_n)\ket{\text{GS}} + \ket{1}_\text{anc}\otimes U^\dag_\text{GS}(L_n - K_n)\ket{\text{GS}}\right)
\end{equation}
We then measure not only $\braket{Z}_\text{anc}$ of the ancillary qubit, but also the expectation value of $\ket{0}\bra{0}$ for each system qubit to obtain the required expectation value.

\section{Quantum Circuit Transpilation}
\label{app:qct}
\begin{figure}[h]
\begin{tikzpicture}
\node[scale=0.7] {
\begin{quantikz}[column sep=4pt]
\lstick{$q_{2,\uparrow}$} & \gate[1]{R_Z} & \gate[1]{\sqrt{X}} & \gate[1]{R_Z} & \gate[2]{C_Z} & \gate[1]{\sqrt{X}} & \gate[1]{R_Z} & \gate[1]{\sqrt{X}} & \qw & \gate[2]{C_Z} & \gate[1]{R_Z} & \gate[1]{\sqrt{X}} & \gate[1]{R_Z} & \qw & \qw & \qw & \qw & \qw & \qw & \qw & \qw & \qw & \qw & \qw & \qw & \qw\\
\lstick{$q_{1,\uparrow}$} & \gate[1]{R_Z}& \gate[1]{\sqrt{X}} & \qw & \qw & \gate[1]{R_Z} & \gate[1]{\sqrt{X}} & \gate[1]{R_Z} & \gate[1]{\sqrt{X}} & \qw & \gate[1]{R_Z} & \gate[1]{\sqrt{X}} & \gate[1]{R_Z} & \gate[1]{\sqrt{X}} & \gate[2]{C_Z} & \gate[1]{R_Z} & \gate[1]{\sqrt{X}} & \gate[1]{R_Z} & \gate[1]{\sqrt{X}} & \gate[2]{C_Z} & \gate[1]{R_Z} & \gate[1]{\sqrt{X}} & \gate[1]{R_Z} \\
\lstick{$q_{0,\uparrow}$} & \gate[1]{R_Z}& \gate[1]{\sqrt{X}} & \gate[2]{C_Z} & \gate[1]{\sqrt{X}} & \gate[1]{R_Z} & \gate{\sqrt{X}} & \gate[2]{C_Z} & \gate[1]{\sqrt{X}} & \gate[1]{R_Z} & \gate[1]{\sqrt{X}} & \gate[1]{R_Z} & \qw & \qw & \qw & \gate[1]{\sqrt{X}} & \gate[1]{R_Z} & \gate[1]{\sqrt{X}} & \qw & \qw & \gate[1]{R_Z} & \gate[1]{\sqrt{X}} & \gate[1]{R_Z}\\
\lstick{$q_{0,\downarrow}$} & \gate[1]{R_Z}& \qw & \qw & \gate[1]{X} & \gate[1]{R_Z} & \qw & \qw & \gate[1]{\sqrt{X}} & \gate{R_Z} & \qw & \qw & \qw & \qw & \gate[2]{C_Z} & \gate[1]{\sqrt{X}} & \gate[1]{R_Z} & \gate[1]{\sqrt{X}} & \qw & \gate[2]{C_Z} & \gate[1]{R_Z} & \gate[1]{\sqrt{X}} & \gate[1]{R_Z}\\
\lstick{$q_{1,\downarrow}$} & \gate[1]{R_Z}& \gate[1]{\sqrt{X}} & \qw &\gate[2]{C_Z} & \gate[1]{R_Z} & \gate[1]{\sqrt{X}} & \gate[1]{R_Z} & \gate[1]{\sqrt{X}} & \gate[2]{C_Z} & \gate[1]{R_Z} & \gate[1]{\sqrt{X}} & \gate[1]{R_Z} & \gate[1]{\sqrt{X}} & \qw & \gate[1]{R_Z} & \gate[1]{\sqrt{X}} & \gate[1]{R_Z} & \gate[1]{\sqrt{X}} & \qw & \gate[1]{R_Z} & \gate[1]{\sqrt{X}} & \gate[1]{R_Z} \\
\lstick{$q_{2,\downarrow}$} & \gate[1]{R_Z}& \gate[1]{\sqrt{X}} & \gate[1]{R_Z} & \qw & \gate[1]{\sqrt{X}} & \gate[1]{R_Z} & \gate[1]{\sqrt{X}} & \qw & \qw & \gate[1]{R_Z} & \gate[1]{\sqrt{X}} & \gate[1]{R_Z} & \qw & \qw & \qw & \qw & \qw & \qw & \qw & \qw & \qw & \qw & \qw & \qw & \qw
\end{quantikz}
};
\end{tikzpicture}
    \caption{Transpiled layer of the Hamiltonian variational ansatz from Fig. \ref{fig:HVA} for the IBM Heron chip (ibm\_torino). Each $R_Z$ gate is parametrized by its own angle.}
    \label{fig:TQC}
\end{figure}
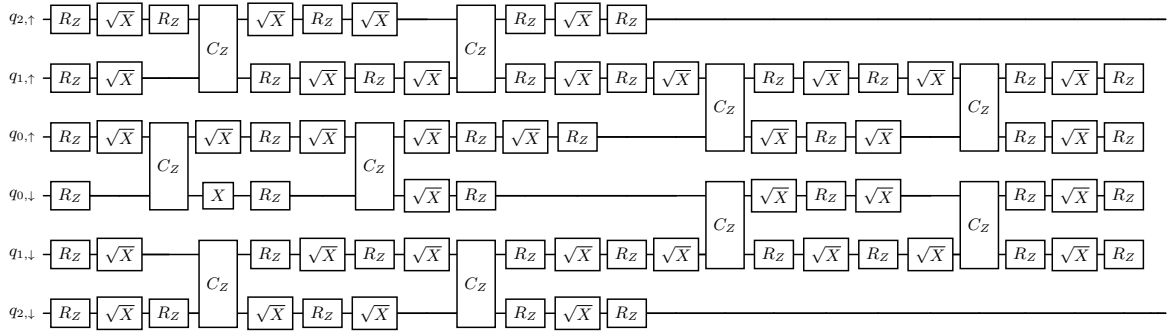

The native gate set of the IBM Heron chip (ibm\_torino) consists of the single-qubit gates
\begin{equation}
    R_Z(\theta) = \left(\begin{array}{cc}
         e^{-\frac{i\theta}{2}} & 0 \\
         0 & e^{\frac{i\theta}{2}} 
    \end{array}\right), \quad 
    X = \left(\begin{array}{cc}
         0 & 1 \\
         1 & 0
    \end{array}\right), \quad 
    \sqrt{X} = \frac{1}{2}\left(\begin{array}{cc}
         1 + i & 1 - i  \\
         1 - i & 1 + i 
    \end{array}\right), \quad \quad
\end{equation}
and the two-qubit gate
\begin{equation}
    C_Z = \left(\begin{array}{cccc}
         1 & 0 & 0 & 0 \\
         0 & 1 & 0 & 0 \\
         0 & 0 & 1 & 0 \\
         0 & 0 & 0 & -1
    \end{array} \right).
\end{equation}
We used IBM's built-in transpiler, including its circuit optimization, to translate our circuits into this native gate set with optimization level 3.
For each layer of our Hamiltonian variational ansatz for the ground state preparation and for the time evolution approximation (compare Fig. \ref{fig:HVA}), the number of two-qubit gates increases from five to ten in the native gate set (see Fig. \ref{fig:TQC}).

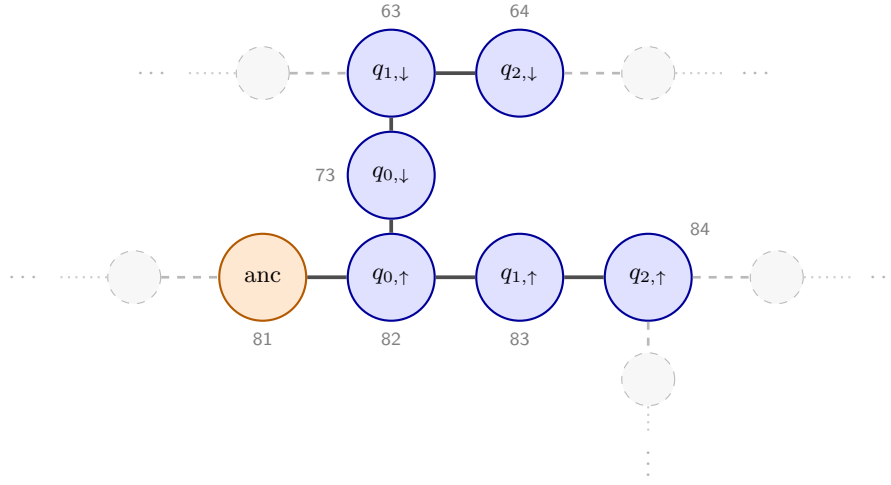
\begin{figure}[h]
\begin{tikzpicture}[
    x=1.7cm, y=1.35cm,
    qubit/.style={circle, draw=blue!60!black, fill=blue!12, line width=0.8pt,
                  minimum size=1.15cm, inner sep=0pt, font=\small},
    anc/.style={qubit, draw=orange!70!black, fill=orange!18},
    ghost/.style={circle, draw=gray!55, dashed, fill=gray!6,
                  minimum size=0.7cm, inner sep=0pt},
    coupler/.style={line width=1.4pt, draw=black!70},
    outer/.style={line width=1pt, draw=gray!55, dashed},
    stub/.style={line width=0.8pt, draw=gray!40, dotted},
    phys/.style={font=\scriptsize\sffamily, text=gray},
  ]
 
  \node[qubit] (63) at (6, 0)  {$q_{1,\downarrow}$};
  \node[qubit] (64) at (7, 0)  {$q_{2,\downarrow}$};
  \node[qubit] (73) at (6,-1)  {$q_{0,\downarrow}$};
  \node[anc]   (81) at (5,-2)  {anc};
  \node[qubit] (82) at (6,-2)  {$q_{0,\uparrow}$};
  \node[qubit] (83) at (7,-2)  {$q_{1,\uparrow}$};
  \node[qubit] (84) at (8,-2)  {$q_{2,\uparrow}$};
 
  \foreach \q/\d in {63/above, 64/above, 73/left, 81/below, 82/below, 83/below, 84/above right}
    \node[phys, \d=1pt of \q] {\q};
 
  \node[ghost] (62) at (5, 0)  {};
  \node[ghost] (65) at (8, 0)  {};
  \node[ghost] (80) at (4,-2)  {};
  \node[ghost] (85) at (9,-2)  {};
  \node[ghost] (93) at (8,-3)  {};
 
  \draw[coupler] (63) -- (64);
  \draw[coupler] (63) -- (73);
  \draw[coupler] (73) -- (82);
  \draw[coupler] (81) -- (82);
  \draw[coupler] (82) -- (83);
  \draw[coupler] (83) -- (84);
 
  \draw[outer] (62) -- (63);
  \draw[outer] (64) -- (65);
  \draw[outer] (80) -- (81);
  \draw[outer] (84) -- (85);
  \draw[outer] (84) -- (93);
 
  \draw[stub] (62) -- ++(-0.6, 0);
  \draw[stub] (65) -- ++( 0.6, 0);
  \draw[stub] (80) -- ++(-0.6, 0);
  \draw[stub] (85) -- ++( 0.6, 0);
  \draw[stub] (93) -- ++( 0,-0.55);
  \node[text=gray!60] (d1) at (4.15, 0)  {$\cdots$};
  \node[text=gray!60] (d2) at (8.85, 0)  {$\cdots$};
  \node[text=gray!60] (d3) at (3.15,-2)  {$\cdots$};
  \node[text=gray!60] (d4) at (9.85,-2)  {$\cdots$};
  \node[text=gray!60] (d5) at (8,-3.75)  {$\vdots$};
 
\end{tikzpicture}
\caption{An example qubit setup on the IBM Heron chip (ibm\_torino). The actual chosen qubits depend on the calibration and fidelity of each qubit on the day of the experiment. With this setup, no fermionic SWAP gates are required, which reduces the circuit depth.}
\label{fig:chip}
\end{figure}

Using the Jordan--Wigner transformation, the impurity and bath sites are mapped to a one-dimensional chain of qubits. 
Since two-qubit gates can only act on physically adjacent qubits, we choose the qubits on the chip as illustrated in Fig. \ref{fig:chip}.
In this layout, each impurity qubit is adjacent to the other impurity qubit and to its first bath qubit, and $q_{0,\uparrow}$ is additionally adjacent to the ancilla qubit required for the Hadamard test. 
This setup does not require any fermionic SWAP gates for implementing any interaction.

\section{Postprocessing}
\label{app:postprocessing}
\begin{figure}[h]
    \centering
    \includegraphics[width=1\linewidth]{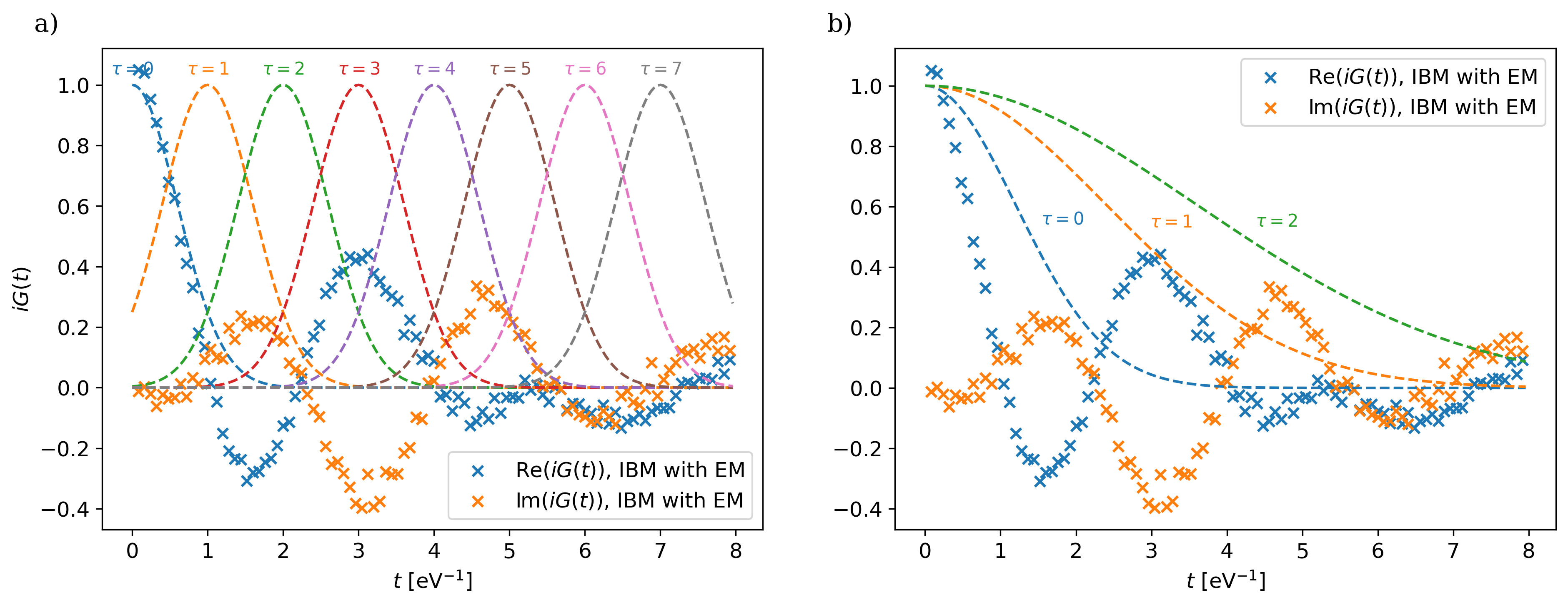}
    \caption{The two window functions Eq. \ref{eq:window1} and Eq. \ref{eq:window2} illustrated for $\tau \in \{0,1,2,3,4,5,6,7\}$ (a) and for $\tau \in \{0,1,2\}$~(b).}
    \label{fig:windowFunctions}
\end{figure}

\begin{figure}[h]
    \centering
    \includegraphics[width = 0.6\linewidth]{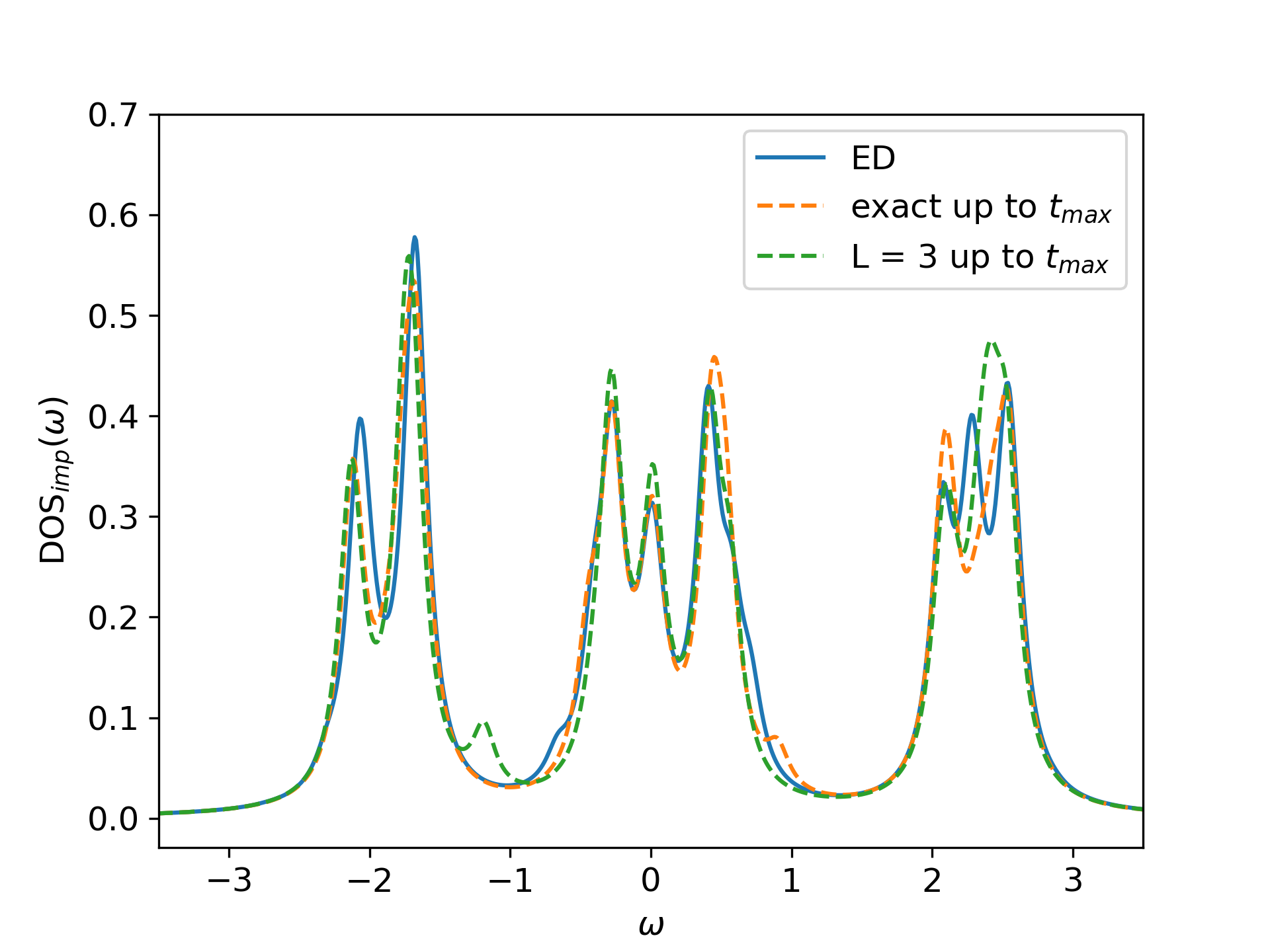}
    \caption{Calculated density of states for the impurity model obtained from exact diagonalization without time restriction (ED), from the exact $G(t)$ restricted to $t\leq t_\text{max}$, and from simulation with our circuit with $L=3$ layers up to $t_{max}$. The latter two are obtained using our described fitting procedure. Here, we used $\eta = 0.1$.}
    \label{app:fitSimulated}
\end{figure}

As stated in the main portion of the article, the reachable time is even with the compressed circuit not sufficient to resolve the energy peaks with the Fourier transformation. 
We therefore use a simple fitting procedure to obtain some of the Lehmann parameters, where we use a short-time Fourier transformation to obtain a good initial guess.
For the window function $w(t - \tau)$ we choose
\begin{equation}
    w_1(t-\tau) = e^{-1.4 \left(t - \tau\right)^2}
    \label{eq:window1}
\end{equation}
with $\tau \in \{0, 1,2,3,4,5,6,7\}$.
Additionally, we used a second window function,
\begin{equation}
    w_2(t-\tau) = e^{-0.35 \left(\frac{t}{\tau + 1}\right)^2},
    \label{eq:window2}
\end{equation}
with $\tau \in \{0, 1, 2\}$, similar to the damping introduced by $\eta$ (see Fig. \ref{fig:windowFunctions}) to increase the number of initial parameters even further.
As illustrated in Fig. \ref{app:fitSimulated}, this method is capable of providing an accurate density of states for exact and simulated results with $L=3$ up to $t_\text{max}$ for a large $\eta$ of $0.1$.
Although the simulated $G(t)$ for $L=3$ differs only slightly from the exact results, additional peaks appear incorrectly.
A more detailed spectral function would still require a simulation up to longer time scales.

\section{Integration of the Quasiparticle Peak}
\label{app:qpw}
\begin{figure}[h]
    \centering
    \includegraphics[width=0.6\linewidth]{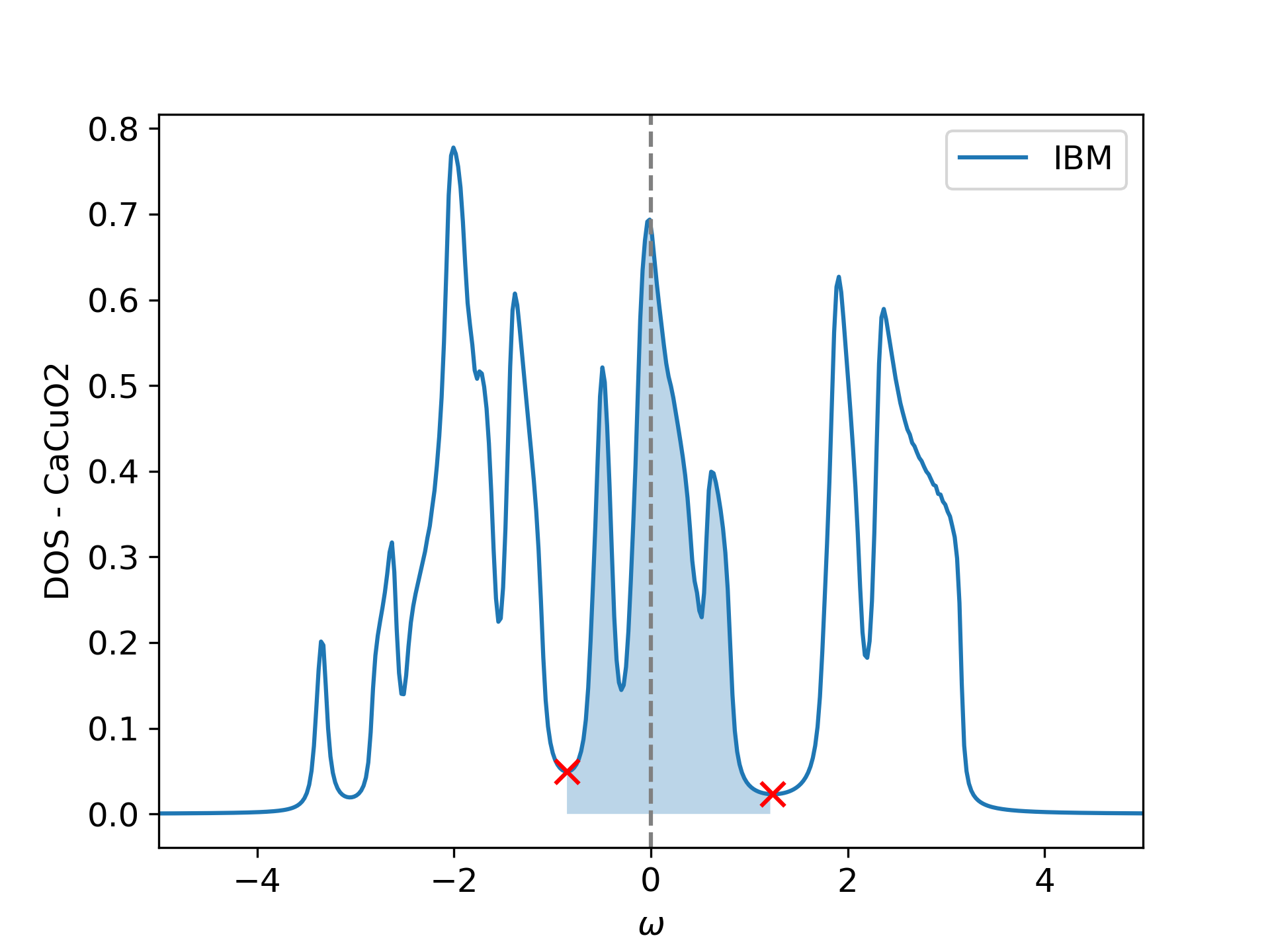}
    \caption{Density of states for \ce{CaCuO2} obtained with the IBM device. We integrate the DOS between the two minima (red crosses) to calculate the quasiparticle weight from the quasiparticle peak (shaded area).}
    \label{fig:appdos}
\end{figure}

Since the signal suffers from noise, the calculation of the quasiparticle weight using Eq. \ref{eq:qpw} is numerically unstable.
We therefore opt to integrate over the central quasiparticle peak of the density of states, similar to Ref. \cite{Keen}, and compare the results using the signal obtained from hardware with exact diagonalization (see Fig. \ref{fig:appdos}). 
The central peak, as we define it here, is the section between the two minima that separate it from the peaks of the lower and upper Hubbard bands. 
This method does not allow for the actual correct value but rather for a qualitative estimate of the actual quasiparticle weight. 

\end{appendix}
\end{document}